\documentclass[fleqn,twoside]{article}
\usepackage{espcrc2,epsfig}

\usepackage{graphicx}
\usepackage[figuresright]{rotating}

\newcommand{\pio}{\pi^0}
\newcommand{\ko}{K^0}

\newcommand{\qcth}{\cos\!\Theta}

\newcommand{\ptau}{{\cal P}_{_{\scriptstyle \!\tau}}}

\newcommand{\aele}{{\cal A}_{_{\scriptstyle \!e}}}
\newcommand{\atau}{{\cal A}_{_{\scriptstyle \!\tau}}}

\newcommand{\alep}{{\cal A}_{_{\scriptstyle \!l}}}

\def\mtau{m_\tau}
\def\mtau2{m^2_{\tau}}
\def\sinw{\sin^2\theta_W}
\def\ee{ \mbox{\rm e}^+ \mbox{\rm e}^-}
\def\tt{ \tau^+\tau^-}

\newcommand{\aone}{\mbox{\rm a}_1}
\def\a1nu{\aone\nu}

\def\s2thw{\sin^{2}\theta^{\mbox{\scriptsize lept}}_{\mbox{\scriptsize eff}}}

\newcommand{\MZ}{M_{\mbox{\footnotesize Z}}}
\newcommand{\TPI}{\tau \rightarrow \pi^\pm \nu}

\newcommand{\TMU}{\tau \rightarrow \mu \nu \bar{\nu}}
\newcommand{\TEL}{\tau \rightarrow {\rm e} \nu \bar{\nu}}

\newcommand{\TRO}{\tau \rightarrow \pi^\pm \pio \nu}
\newcommand{\TAA}{\tau \rightarrow \pi^\pm 2 \pio \nu}
\newcommand{\TPMP}{\tau \rightarrow h^\pm 3 \pio \nu}
\newcommand{\TPMM}{\tau \rightarrow h^\pm 4 \pio \nu}
\newcommand{\TPPP}{\tau \rightarrow 3 \pi^\pm \nu_{\tau}}
\newcommand{\TPPPPP}{\tau \rightarrow 5 \pi^\pm \nu_{\tau}}
\newcommand{\TPPPO}{\tau \rightarrow 3 \pi^\pm \pio \nu_{\tau}}
\newcommand{\TPPPPPO}{\tau \rightarrow 3 \pi^\pm \pio \nu_{\tau}}
\newcommand{\TPPPOO}{\tau \rightarrow 3 h^\pm 2 \pio  \nu_{\tau}}

\newcommand{\afb}{A_{FB}}
\newcommand{\ecm}{E_{cm}}

\newcommand{\AmS}{{\protect\the\textfont2
  A\kern-.1667em\lower.5ex\hbox{M}\kern-.125emS}}

\title{$\tau$ physics at LEP}

\author{Francisco Matorras\address[IFCA]{Instituto de F\'{\i}sica de Cantabria\\
	Santander, Spain}}
       
\begin{document}

\begin{abstract}
This paper gives an overview of some of the more interesting results obtained 
	at LEP in $\tau$ 
	physics: precision measurements in 
	neutral and charged currents universality and structure,
	$\tau$  mass, topological and exclusive Branching Ratios. \vspace{1pc}
\end{abstract}

\maketitle

\section{Introduction}
The production and decay of the $\tau$ lepton at LEP has provided a large variety
of tests of the Standard Model.
At LEP I, the first phase of the collider running at centre of mass energies close to $M_Z$, the $\tau$ are predominantly
produced in pairs, from the annihilation of an electron-positron pair into a Z boson. In total, 
about 500000 pairs
have been collected by the four LEP experiments, providing a unique environment to study the couplings of the
Z to heavy leptons, where many extensions of the Standard Model would induce significant modifications to the
expectations.

At these energies, more than 20 charged particles are produced on average in hadronic events, allowing a very 
efficient and clean separation from the tau pairs.
This fact, together with the significant amount of recorded $\tau$, 
allows further precision studies on its decay: the leptonic
decays are a very noticeable place to study the charged current
universality and structure; the hadronic decays allow the study of the strong interaction at an intermediate
$q^2$; neutrino physics...

At LEP II, where the centre of mass energy has been progressively increased until 208 GeV, two new broad
subjects were opened: the study of the $\tau$ production in the decay of a $W$, permitting additional tests on the
charged current and new particles searches, where the $\tau$ plays an interesting role. Each experiment 
recorded about 700 $pb^{-1}$, which represented about 50000 W pairs produced in total. 

Only some these subjects will be reviewed here, more details being available in other talks of this
conference~\cite{davier,gui,davide}.

\section{Neutral currents}
Different observables in the production of tau pairs through the decay of a Z boson, 
produced in a $\ee$ annihilation at different energies allow the precise measurement of
the weak neutral couplings and as a consequence indirect bounds on new physics are set.
The structure of the neutral currents, weak and electromagnetic, was also
studied.

\subsection{Lineshape at the Z resonance}
\begin{figure}[htbp]
\vspace{-1cm}
\begin{center}
\mbox{\epsfxsize 8.0cm\epsfbox{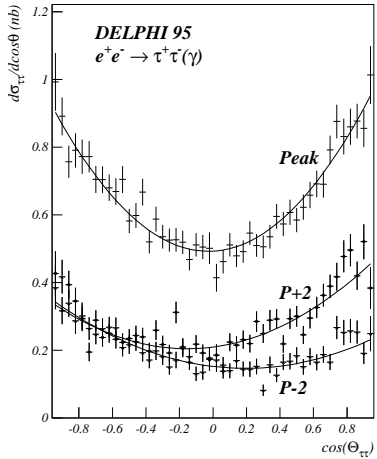}}
\end{center}

\vspace{-2.0cm}
\caption
{Forward-backward charge asymmetry. Polar angle distribution for tau pairs 
measured by the DELPHI experiment compared to the fit result.}
\label{fig:asim1}
\end{figure}

\begin{figure}[htbp]
\vspace{-1cm}
\begin{center}
\mbox{\epsfxsize 8.0cm\epsfbox{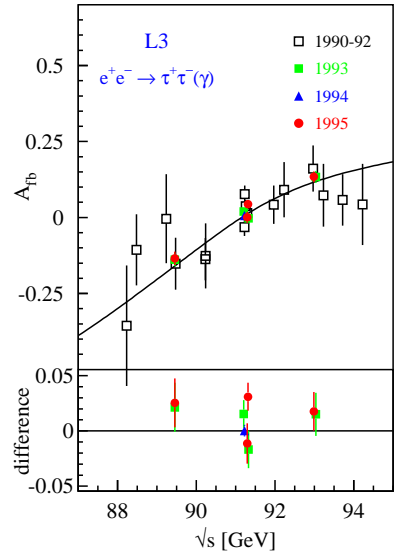}}
\end{center}
\vspace{-1.5cm}
\caption
{Centre of mass energy dependence of $A_{FB}$ measured by the L3 experiment. 
compared to the Standard Model prediction. 
The bottom part of the plot shows the residual difference between the
measured points and the expectation.}
\label{fig:asim2}
\end{figure}

The production cross section was extensively studied by all LEP experiments, as a function of the centre-of-mass energy.
Results are often expressed in terms of the width ($\Gamma_\tau$=$\Gamma(Z\to\tau\tau$)) or of its ratio
to the hadronic width: $R_\tau=\Gamma_{had}/\Gamma_\tau$.
$\Gamma_\tau$ is proportional to ${g_a}^2+{g_v}^2$, the axial and vectorial couplings to the Z. 
Being the vectorial coupling much smaller than the axial, this observable essentially provides 
a measurement of the axial coupling.
Another observable studied at LEP is the forward-backward charge asymmetry, $\afb$, defined as the
fraction of $\tau^-$ produced in the forward direction, defined according to the incoming electron.
This quantity, at tree level, at $\ecm=\MZ$ and neglecting $\gamma$ exchange, can be expressed as
$\afb=\aele \atau$, where $\alep=\frac{2{g_v}{g_a}}{{g_a}^2+{g_v}^2}\approx2\frac{g_v}{g_a}$,
providing information on the vector coupling, but not on its sign. 
An example of the angular dependence used to extract the asymmetry is shown in
figure~\ref{fig:asim1}. Figure~\ref{fig:asim2} shows the dependence of this magnitude with the centre of mass
energy.

\begin{figure}[htb]
\begin{center}
\mbox{\epsfxsize 8.0cm\epsfbox{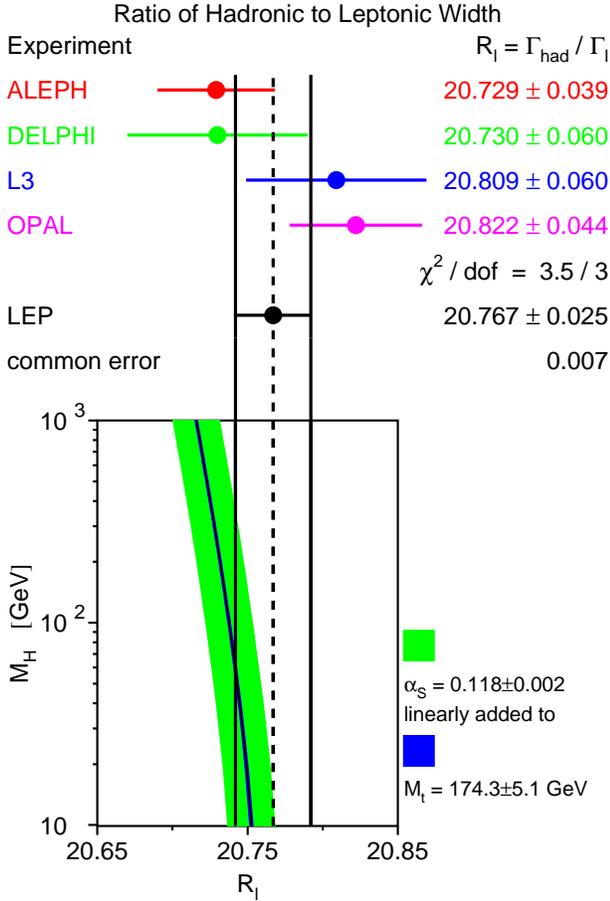}}
\end{center}
\vspace{-1.0cm}
\caption
{LEP results for the leptonic width of the Z, 
compared to the Standard Model expectation (see~\cite{lepeew} for more details).}
\label{fig:lepgam}
\end{figure}

\begin{figure}[htb]
\begin{center}
\mbox{\epsfxsize 8.0cm\epsfbox{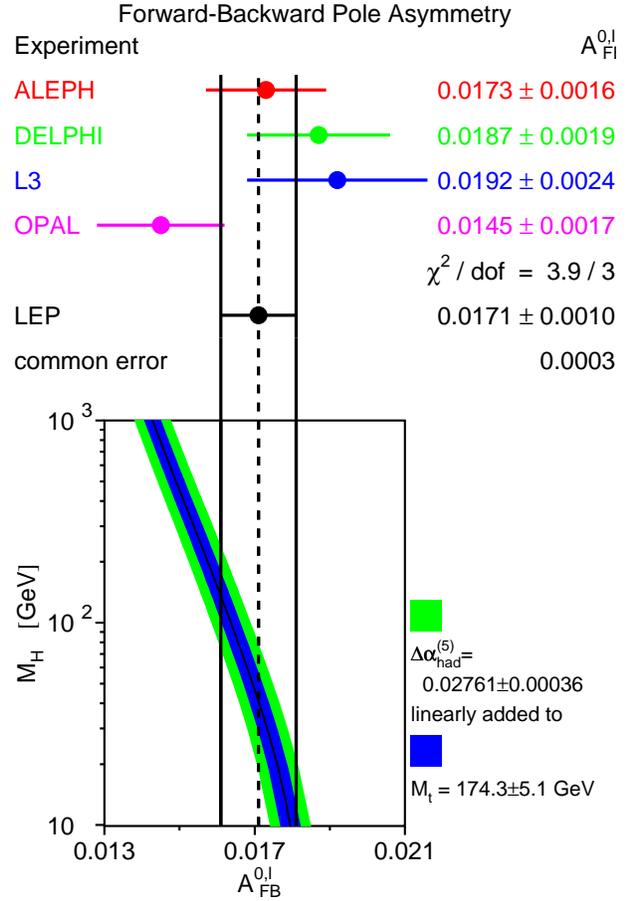}}
\end{center}
\vspace{-1.0cm}
\caption
{LEP results for the  forward backward charge asymmetry, 
compared to the Standard Model expectation (see~\cite{lepeew} for more details).}
\label{fig:lepafb}
\end{figure}

The final results for each of the LEP collaborations and its combination, together with the
comparison with the Standard Model prediction as a function of the top quark and Higgs
boson masses  are shown in figures~\ref{fig:lepgam} and~\ref{fig:lepafb} 
(more details in~\cite{lepeew}).


\subsection{$\tau$ polarisation}
\begin{figure}[htbp]
\hspace{-0.5cm}
\mbox{\epsfysize 14.0cm\epsfxsize 8.0cm\epsfbox{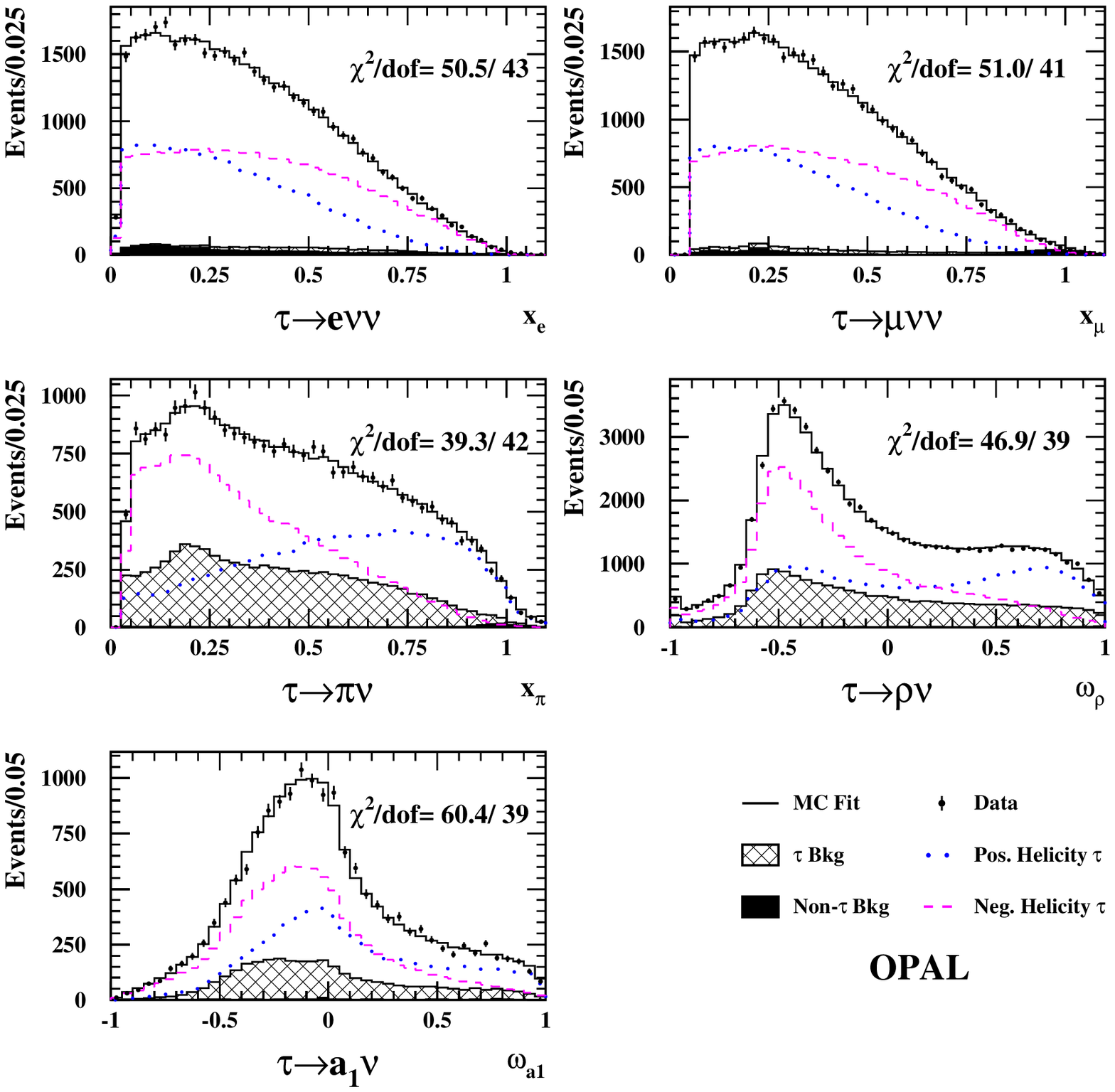}}
\vspace{-1.0cm}
\caption
{Distribution of the polarisation estimators for the different channels studied
by the OPAL experiment, compared with the Standard Model
expectation for positive or negative helicity.}
\label{fig:ptau}
\end{figure}

The fermions pairs produced in the annihilation of a Z are polarised. This polarisation is only
measurable for $\tt$, because the $\tau$ lepton decays inside the detector through a weak interaction,
violating parity. The $\tau$ spin can be inferred from its decay products.
Different polarisation estimators~\cite{rouge} can be built from the measured 
kinematical quantities, once the decay channel is identified. 
Figure~\ref{fig:ptau}
shows an example of the different distributions used and its sensitivity to the
polarisation.

\begin{figure}[htbp]
\begin{center}
\mbox{\epsfxsize 8.0cm\epsfbox{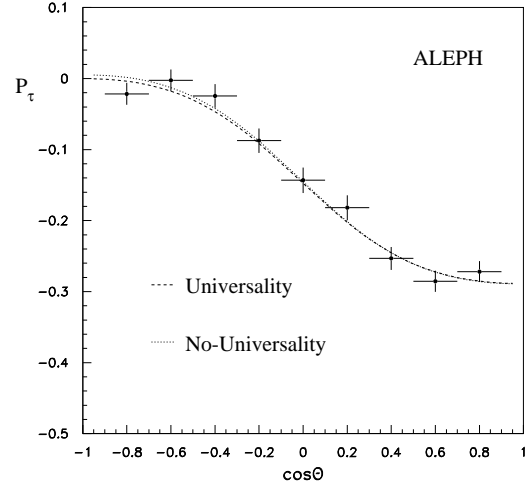}}
\end{center}
\vspace{-1.0cm}
\caption
{Angular dependence of $\ptau$ as measured by the ALEPH experiment, compared with the Standard Model
expectation.}
\label{fig:ptaucos}
\end{figure}

Furthermore, this polarisation depends on the polar angle through the expression (at tree level):
\begin{equation}
  \ptau(\qcth) =
  \frac{\aele \cdot (1+\cos^2 \!\Theta) + \atau \cdot 2\qcth}
       {(1+\cos^2 \!\Theta) + \frac{4}{3} \afb \cdot 2\qcth}.
\label{eqn:polcos0}
\end{equation}
This dependence was studied by the four experiments, measuring the average
polarisation in restricted polar angle ranges.
An example of the result is shown in Figure~\ref{fig:ptaucos}.
The measurement of the polarisation as a function of the production polar angle provides an almost
independent measurement of $\aele$ and $\atau$, and as a consequence, independent measurement on the
$\tau$ and electron vector coupling (including its sign). 

\begin{figure}[htb]
\begin{center}
\mbox{\epsfxsize 8.0cm\epsfbox{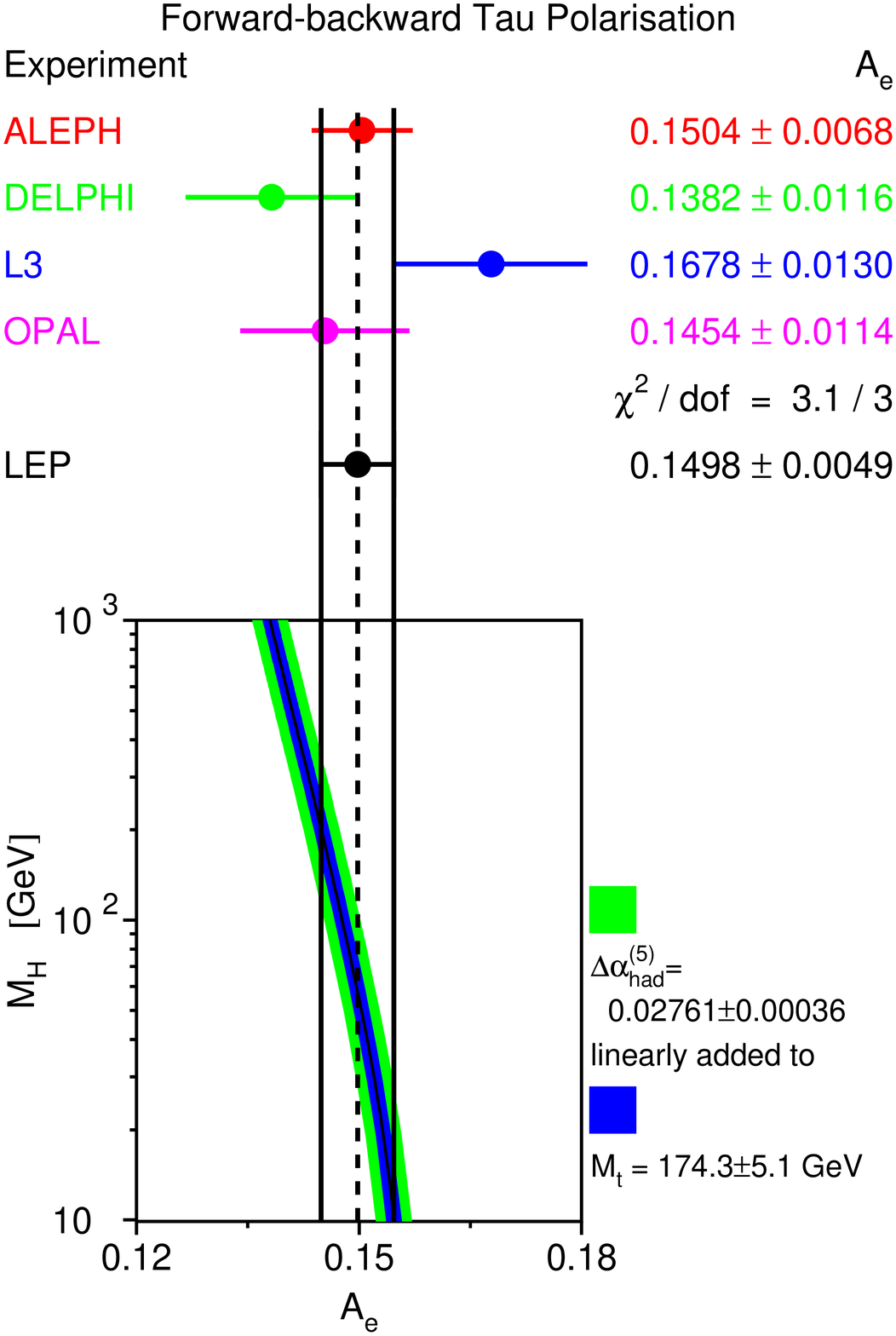}}
\end{center}
\vspace{-1.0cm}
\caption
{LEP results for $\aele$ obtained from the tau polarisation, compared to the Standard Model expectation
(see~\cite{lepeew}).}
\label{fig:aele}
\end{figure}

\begin{figure}[htb]
\begin{center}
\mbox{\epsfxsize 8.0cm\epsfbox{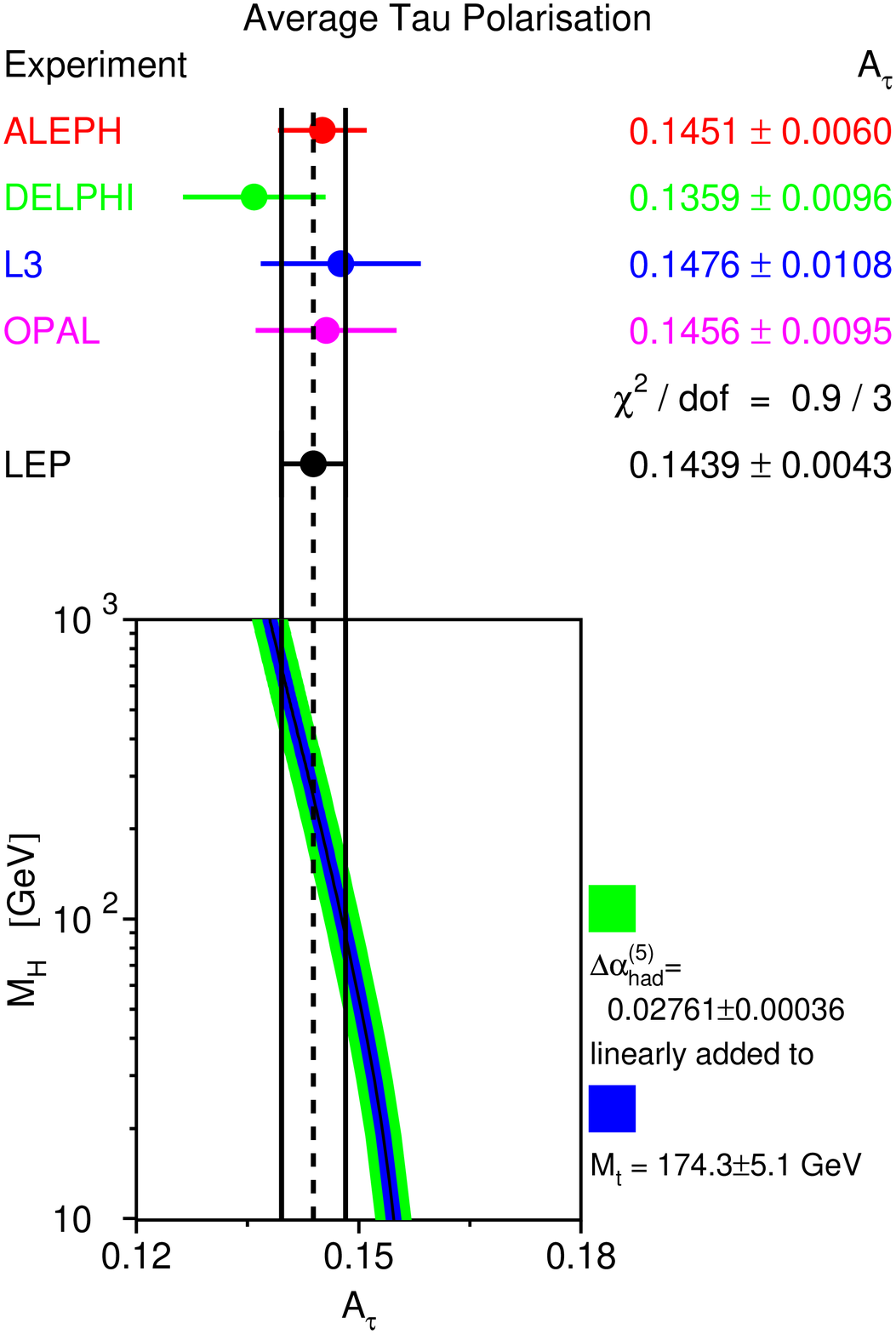}}
\end{center}
\vspace{-1.0cm}
\caption
{LEP results for $\atau$ obtained from the tau polarisation, compared to the Standard Model expectation
(see~\cite{lepeew}).}
\label{fig:atau}
\end{figure}

The final results are
shown in figure~\ref{fig:aele} and ~\ref{fig:atau}, together with its combination accounting for
common systematic errors\cite{lepeew}.


\subsection{Universality}
\begin{figure}[htbp]
\begin{center}
\mbox{\epsfxsize 8.0cm\epsfbox{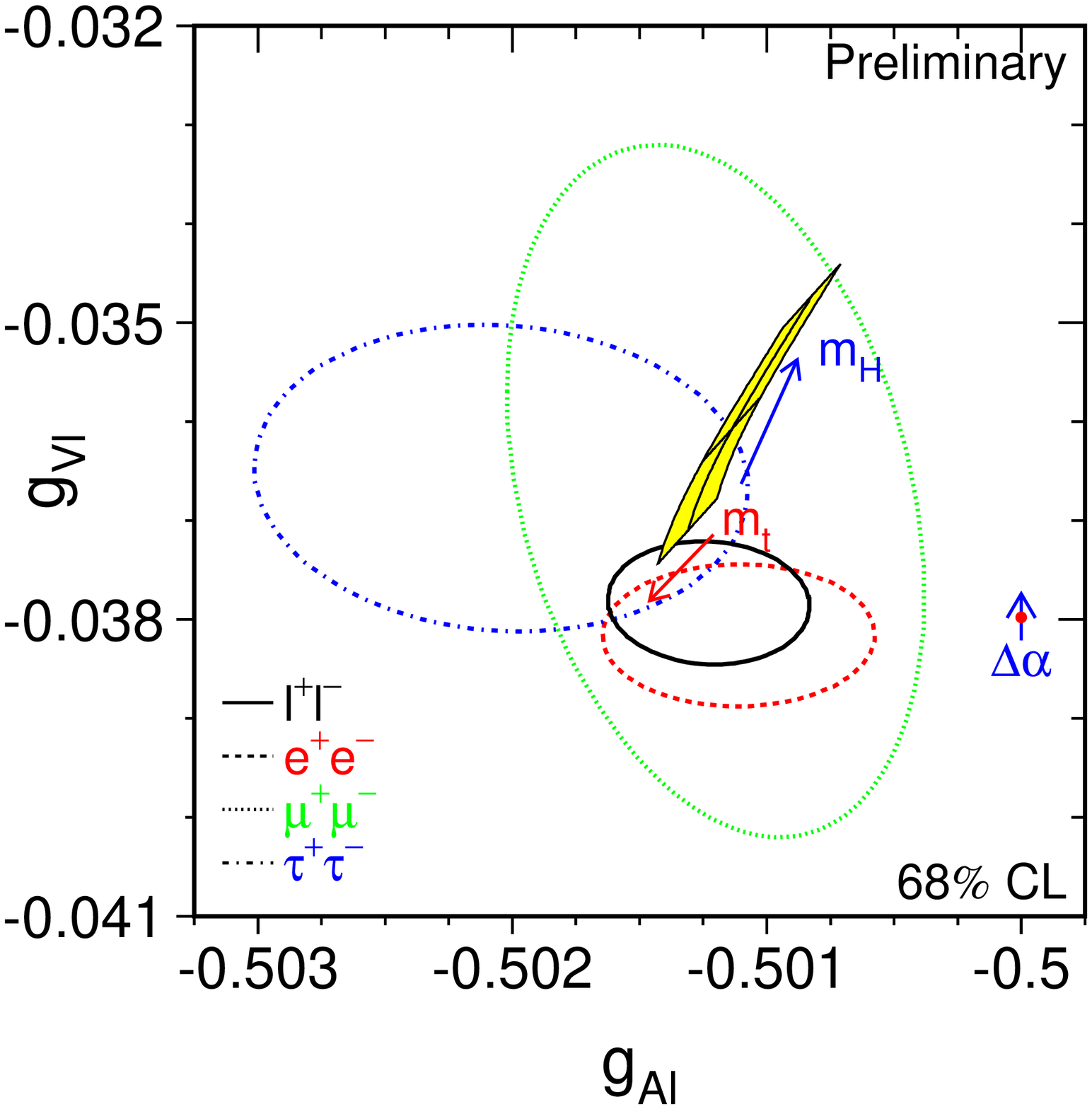}}
\end{center}
\vspace{-1.6cm}
\caption
{Results on the leptonic vector and axial vector couplings to the Z, obtained from
the fit of all observables . The different lines show the 2D contours corresponding to 68\% CL, for each of the
leptons and for their combination.}
\label{fig:gvga}
\end{figure}

\begin{figure}[htbp]
\vspace{0.2cm}
\begin{center}
\mbox{\epsfxsize 6.5cm\epsfbox{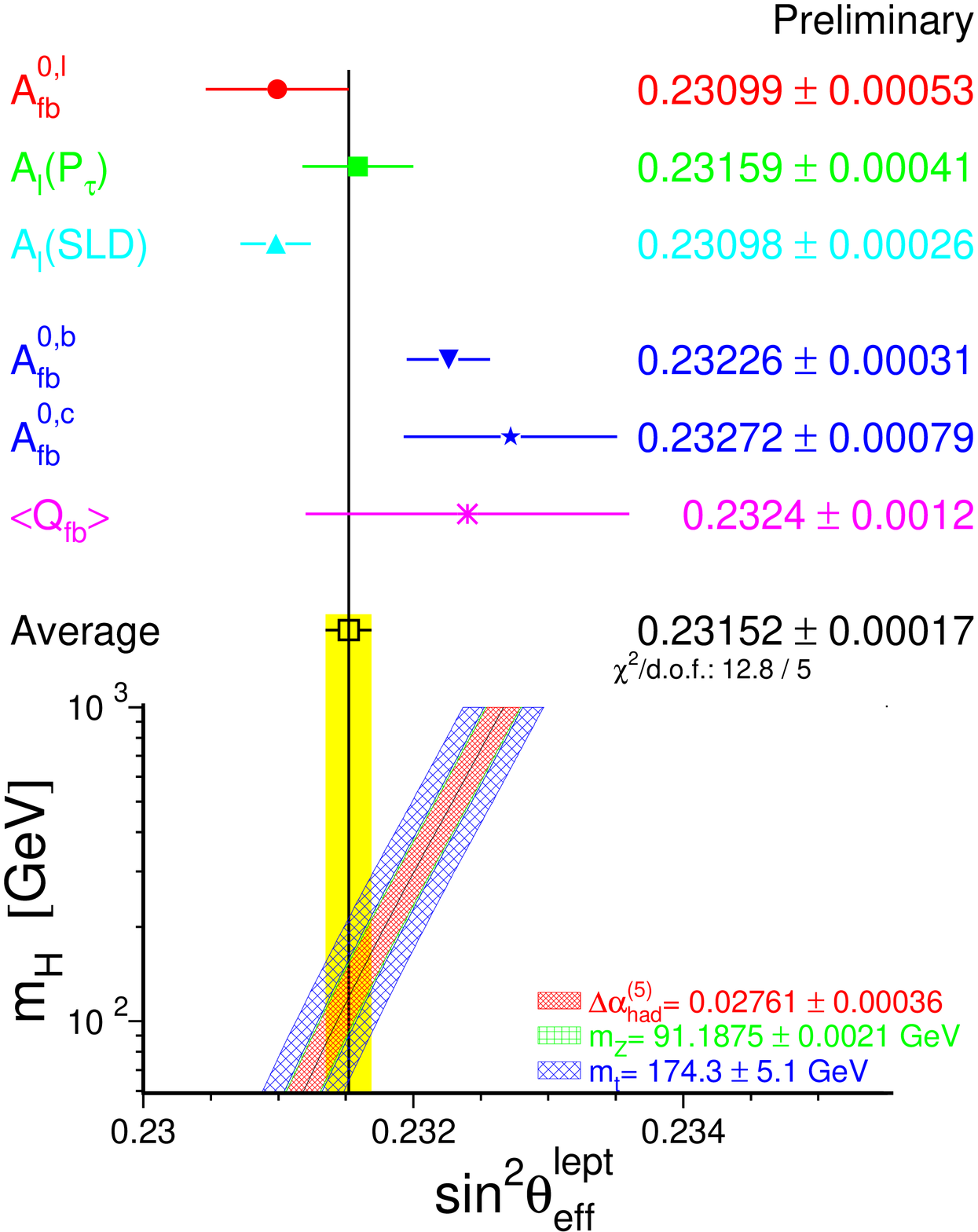}}
\end{center}
\vspace{-1.0cm}
\caption
{
Comparison of the $\sinw$ weak mixing angle as obtained by different methods at LEP and SLD.}
\label{fig:sinw}
\end{figure}

The measurements discussed above are treated coherently in the Standard Model framework, including radiative
corrections, to extract the vector and axial couplings to the Z to any lepton. The results~\cite{lepeew} are
summarised in figure~\ref{fig:gvga}, where the results from SLD are also included. The agreement between the three
68\% CL surfaces shows that the universality is fulfilled, setting bounds on unknown Standard Model parameters and
new physics (see~\cite{lepeew} for more details). It can also be appreciated that the
inclusion of tau polarisation measurements improves significantly the precision on the axial coupling
to electron and taus.
Accepting that the universality is fulfilled, one can extract the value of $\sinw$ from each of these
measurements. The results are shown in figure~\ref{fig:sinw} and compared to those obtained through other
observables. We can see that the combination of 
tau observables, give a similar precision to the most precise b quark charge asymmetry.

\subsection{Electric and magnetic dipole moments}
The structure of the electromagnetic interaction neglecting radiative corrections would be purely vector.
However, the radiative corrections distort this picture. The Standard Model predicts a magnetic dipole 
($a_\tau^\gamma$)
of 0.11, while  the electric dipole ($d_\tau^\gamma$) is still exactly 0. Models like lepton compositeness or CP
violation scenarios predict a non zero electric dipole and an enhancement of the magnetic dipole.

OPAL~\cite{opaldipoles} and L3~\cite{l3dipoles} use the $E_\gamma$ differential cross section in $\tt\gamma$ events 
at the Z.
The photon energy spectrum is fitted to a linear combination of the
{\bf SM} expectation plus an additional contribution from terms with  
anomalous magnetic or electric dipoles. The limits at 95\% CL are:
\begin{flushleft}
$-0.052<a_\tau^\gamma<0.058$~~(L3)\\
$-0.068<a_\tau^\gamma<0.065$~~(OPAL)\\
$-3.1~10^{-16}<Re(d_\tau^\gamma)<3.1 ~10^{-16}$~~e~cm~~(L3)\\
$-3.8~10^{-16}<Re(d_\tau^\gamma)<3.2 ~10^{-16}$~~e~cm~~(OPAL)\\
\end{flushleft}

DELPHI~\cite{delphigg} and L3~\cite{l3gg} also used the $\gamma\gamma$ collisions at or above the Z. The production 
cross section of this process is also sensitive to the anomalous dipole moments, allowing an additional measurement.
The preliminary results are:
\begin{flushleft}
$-0.062<F_2<0.044$~~(L3)\\
$-0.017<a_\tau^\gamma<0.019$~~(DELPHI)\\
$|Re(d_\tau^\gamma)|<6.7 ~10^{-16}$~~e~cm~~(L3)\\
$|Re(d_\tau^\gamma)|<3.8 ~10^{-16}$~~e~cm~~(DELPHI)\\
\end{flushleft}
where $F_2$ is an average form factor not extrapolated to $q^2=0$, $a_\tau^\gamma=F_2(q^2=0)$.

In addition, different studies interpret many precision measurement of tau observables in the framework of a possible
non-zero dipoles. One of these studies~\cite{gabriel} uses LEP I tau production cross sections, forward-backward
asymmetry and polarisation measurements together with LEP II and Tevatron results on $W$ decay to $\tau$ 
to set the limit $-0.007<a_\tau^\gamma<0.005$ at 95\% CL . A similar
study~\cite{escribano} based only on LEP I data 
sets the limit $-0.004<a_\tau^\gamma<0.006$ and $|Re(d_\tau^\gamma)|<1.1~ 10^{-17}$ \mbox{e cm} at 95\% CL.

Similarly, the radiative corrections modify the pure vector/axial$-$vector structure of the weak
neutral current. Very small values are predicted by the Standard Model, but again there is some enhancement in many
extensions of the Standard Model. 
All LEP experiments have studied the existence of a CP-violating electric-weak dipole using a CP-odd observable,
based on tau spin correlation.
L3 has additionally investigated the weak-magnetic dipole defining several parity odd azimuthal
asymmetries in hadronic decays. No new results have been published in the last two years and therefore the
interested reader is addressed to previous reviews~\cite{duncan}.

\subsection{Lineshape above the Z resonance}
\begin{figure}[htbp]
\begin{center}
\mbox{\epsfxsize 6.0cm\epsfbox{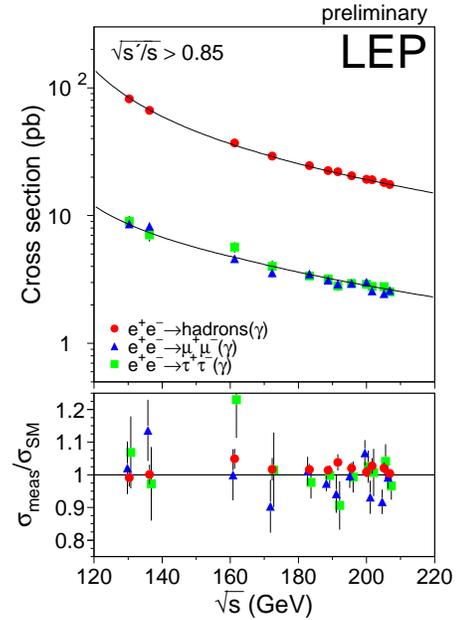}}
\end{center}
\vspace{-1.0cm}
\caption
{Leptonic cross section 
$\ee$ annihilation as a function of the centre of mass energy above the Z resonance.
The bottom part of the plot shows the residual difference between the
measured points and the expectation.}
\label{fig:hesig}
\end{figure}

\begin{figure}[htb]
\begin{center}
\mbox{\epsfxsize 8.0cm\epsfbox{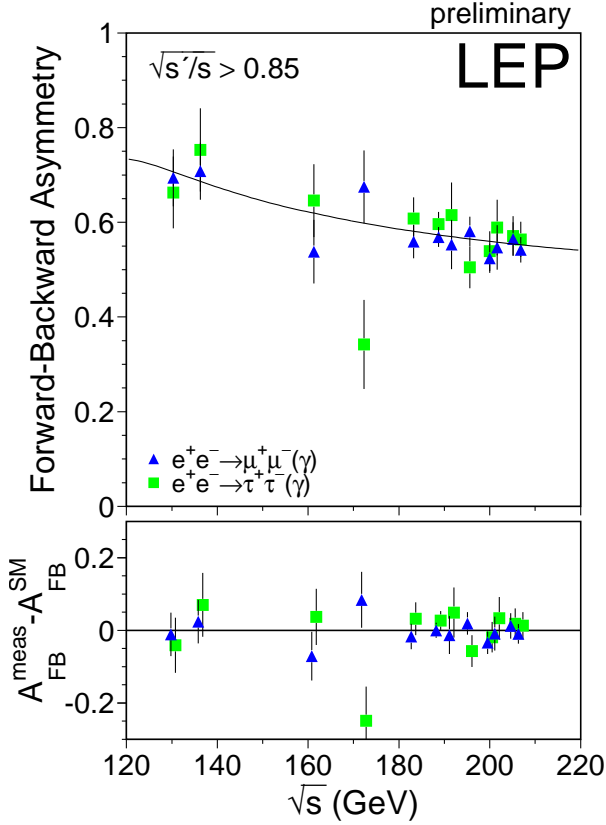}}
\end{center}
\vspace{-1.0cm}
\caption
{Forward-backward  charge asymmetry in 
$\ee$ annihilation as a function of the centre of mass energy above the Z resonance.
The bottom part of the plot shows the residual difference between the
measured points and the expectation.}
\label{fig:heafb}
\end{figure}

\begin{figure}[htb]
\vspace{1cm}
\begin{center}
\mbox{\epsfxsize 8.0cm\epsfbox{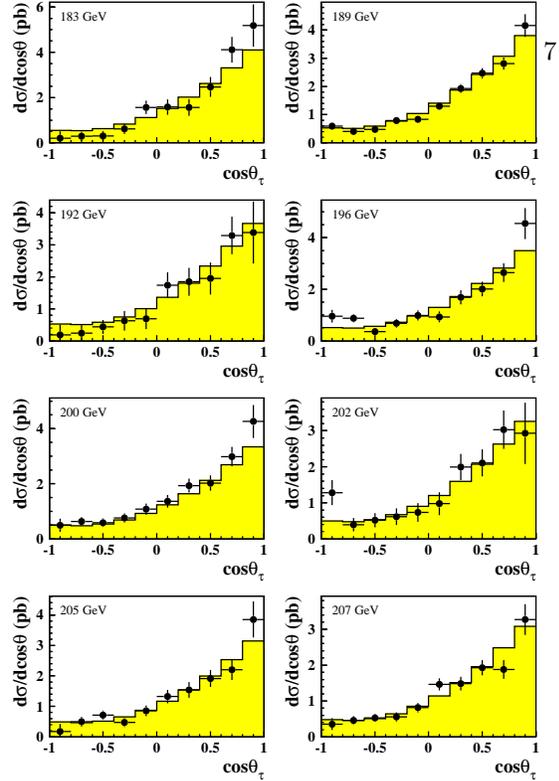}}
\end{center}
\vspace{-1.0cm}
\caption
{$\tt$ differential cross section in 
$\ee$ annihilation as a function of the centre of mass energy above the Z resonance.}
\label{fig:hediff}
\end{figure}

All LEP experiments have extended the lineshape measurement to the higher energy runs up to centre of mass
energy of 208 GeV. At these energies, an additional precision measurement of the $\gamma - Z$ interference is 
performed.  The improvement in the precision of the weak couplings
is marginal, but on the contrary there is a significant sensitivity 
to new physics such as  models with additional Z bosons
or with contact interactions. Results are shown in figures~\ref{fig:hesig} to ~\ref{fig:hediff}~\cite{lepeew}.
Good agreement with the standard Model was found in all cases and therefore the results were used to set limits on new
physics~\cite{lepeew}.
DELPHI has also measured $\ptau$ at higher energies~\cite{igorptau}, giving a result of $\ptau=-0.16\pm0.13\pm0.05$ for an
average centre of mass energy of 190 GeV, consistent with the Standard Model expectation.

\section{Charged currents}
The weak decay of the tau inside the LEP detectors allows a number of interesting
studies of the weak charged current. The W boson decay to leptons provides further
information for equivalent tests, but at a different energy scale and with an `on-shell' W. In
particular, the decay $W\to \tau \nu$ is sensitive to many extensions of the Standard Model.

\subsection{Universality in tau decays}
In the Standard Model and assuming V-A coupling and massless neutrino, 
the tau decay leptonic widths are given by
\begin{equation}
\label{eqn:gammal}
 \Gamma(\tau \rightarrow l \nu_{\tau} \overline{\nu}_{l}) =
 \frac{g_{l \tau}^{2}m_{\tau}^{5}}{192 \pi^3} f(\frac{m_{l}^2}{m_{\tau}^2})
r_{RC}^{\tau}
\end{equation}
 Here, $f(\frac{m_{l}^2}{m_{\tau}^2})$  
 is a phase space
 factor with value $1.0000$ for electrons
 and $ 0.9726 $ for muons.
The quantity $r_{RC}^{\tau}$ is a factor due to electroweak radiative
 corrections, which
 has the value 0.9960, in both leptonic decays. Then, the ratio of the two leptonic widths (or Branching 
 Ratios), provides a direct comparison of the 
 couplings $g_e$ and $g_\mu$.
\begin{equation}
\label{eqn:brrat}
\frac{B(\TMU)} {B(\TEL)} =
\frac{g_{\mu}^{2}} {g_{e}^{2}} \cdot
\frac{f(\frac{m_{\mu}^{2}}{m_{\tau}^{2}})} {f(\frac{m_{e}^{2}}{m_{\tau}^{2}})}
\end{equation}

\begin{figure}[htbp]
\begin{center}
\mbox{\epsfxsize 8.5cm\epsfbox{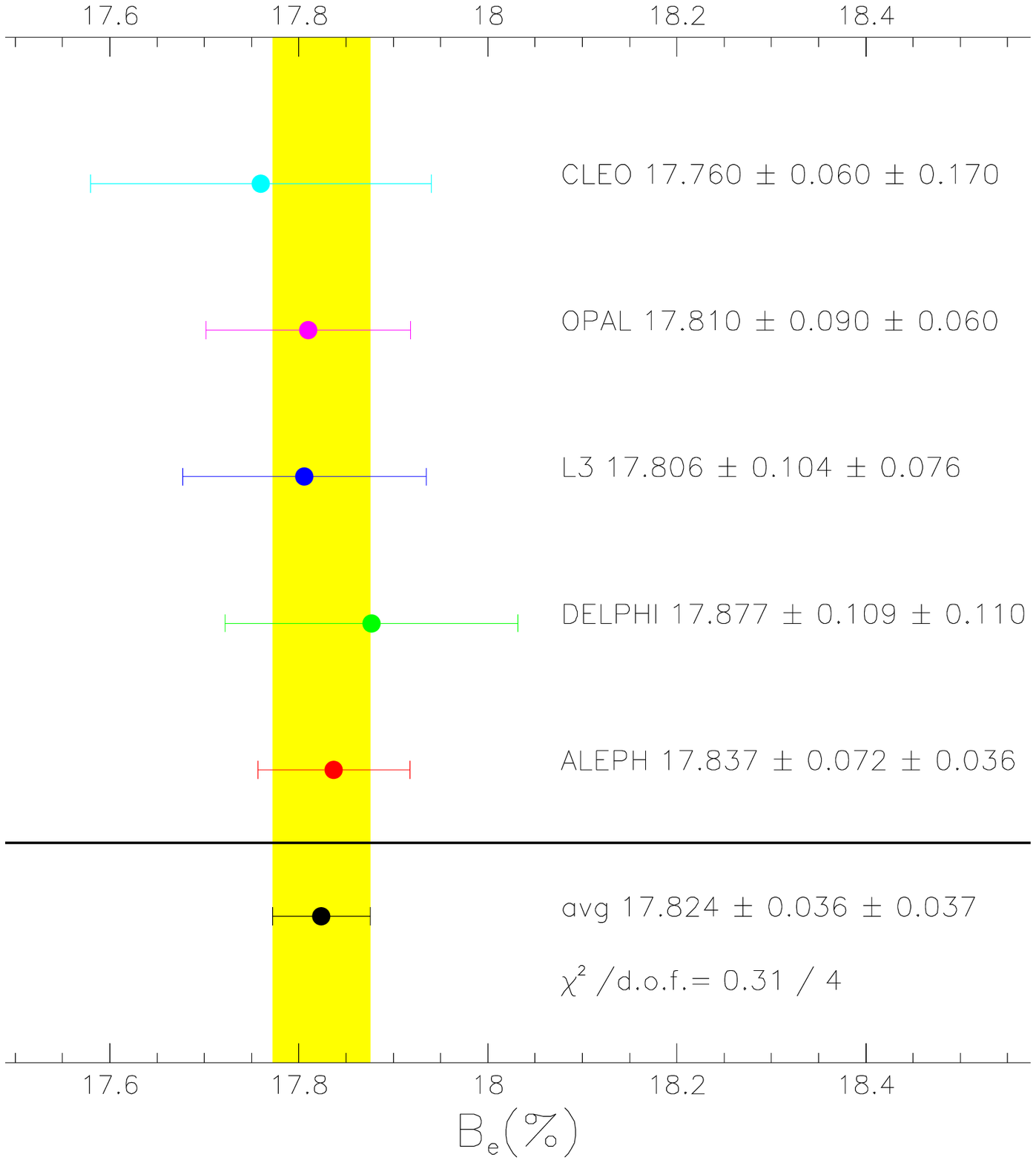}}
\end{center}
\vspace{-1.0cm}
\caption
{Results on the $\TEL$   Branching Ratio.}
\label{fig:bele}
\end{figure}

\begin{figure}[htbp]
\begin{center}
\mbox{\epsfxsize 8.5cm\epsfbox{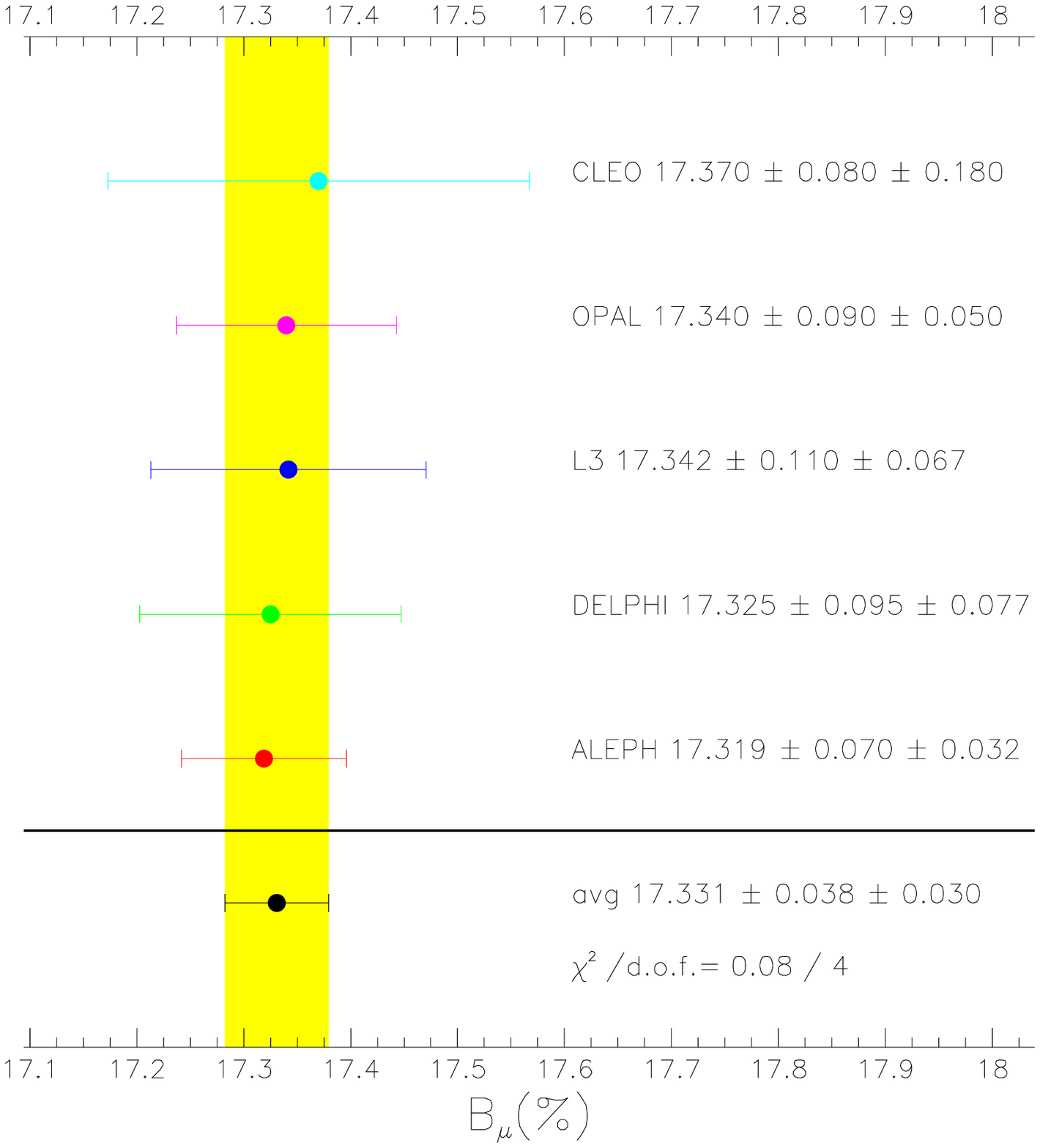}}
\end{center}
\vspace{-1.0cm}
\caption
{Results on the $\TMU$  Branching Ratio.}
\label{fig:bmu}
\end{figure}

ALEPH~\cite{davier} and OPAL~\cite{opalbmu} have recently presented new preliminary results on the leptonic
Branching Ratios, while DELPHI~\cite{delphilep} and L3~\cite{l3lep} have published their final results.
These results are summarised in figures~\ref{fig:bele} and ~\ref{fig:bmu}
together with the results obtained by other
experiments~\cite{pdg} and the averages\footnote{ The averages are done here and thereafter 
 assuming uncorrelated systematic errors between experiments.}. 
The comparison between the two Branching Ratios using expression~\ref{eqn:brrat} yields
$g_\mu/g_e=0.9999\pm0.0020$, perfectly compatible with electron-muon universality. This measurement
is almost as precise as the similar test done in pion decay 
($g_\mu/g_e=1.0017\pm0.0015$).

\begin{figure}[htbp]
\begin{center}
\mbox{\epsfxsize 8.5cm\epsfbox{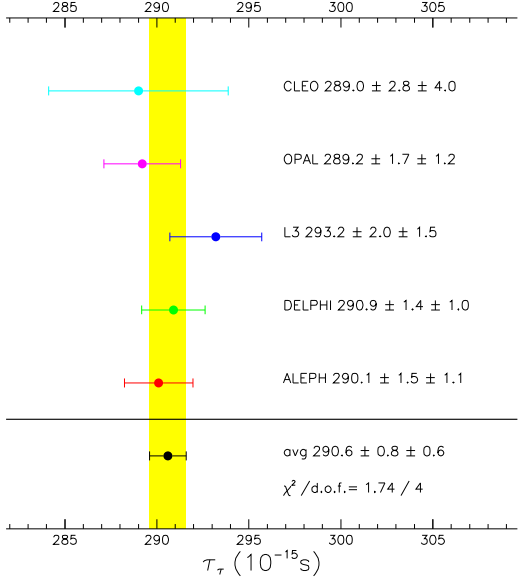}}
\end{center}
\vspace{-1.0cm}
\caption
{Results on the $\tau$ lifetime.}
\label{fig:lifetime}
\end{figure}

Using the $\tau$ and $\mu$ mass and lifetime measurements, together
with the analogue of equation \ref{eqn:gammal} for muon decay, a further
universality test can be performed, comparing the $\tau$ couplings to those of the lighter leptons. 
 Figure~\ref{fig:lifetime} shows the current results on the lifetime~\cite{alephlifetime,l3lifetime,opallifetime}, 
 including new preliminary results
 from DELPHI~\cite{delphilifetime}. Assuming \mbox{$e$ - $\mu$} universality,   
 we can combine the electron and muon Branching Ratios (correcting the second to 
 account for mass effects) into a single BR for a massless lepton and then 
 the comparison of the muon and tau lifetimes
 yields: $g_\tau/g_l=1.0004\pm0.0010(BR)\pm0.0017(\tau_\tau)\pm0.0004(m_\tau)$. The error is split into the
 contributions from the leptonic Branching Ratios, the $\tau$ lifetime and mass, respectively, being other
 contributions negligible.
 
\begin{figure}[htbp]
\begin{center}
\mbox{\epsfxsize 8.5cm\epsfbox{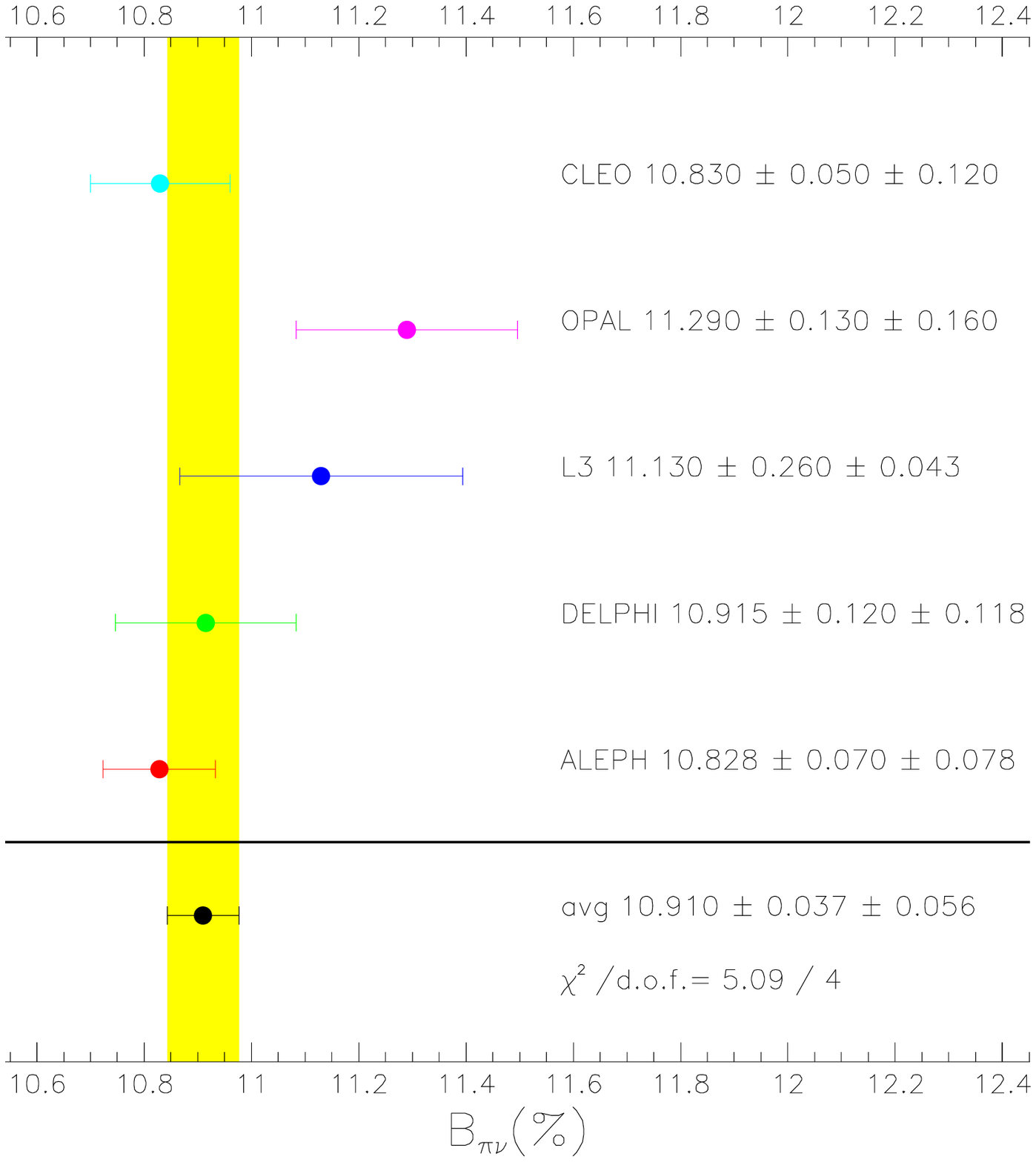}}
\end{center}
\vspace{-1.0cm}
\caption
{Results on the $\TPI$ Branching Ratio.}
\label{fig:bpi}
\end{figure}

There is another possible test comparing the
tau lifetime and the widths $\Gamma(\tau \to \pi\nu)$ and $\Gamma(\tau \to K \nu)$ 
with the $\pi$ and $K$ lifetimes and the widths $\Gamma(\pi \to l\nu)$ and $\Gamma(K \to l\nu)$.
ALEPH~\cite{davier} and DELPHI~\cite{delphihad} presented recently new preliminary measurements of the Branching Ratio to $\TPI$.
Figure~\ref{fig:bpi} shows these results together with those of other experiments.
The comparison gives $g_\tau/g_l=1.0000\pm0.0033(BR)\pm0.0017(\tau_\tau)\pm0.0002(m_\tau)$.

\subsection{W leptonic decays}
\begin{figure}[htb]
\vspace{3.cm}
\hspace{-2.cm}
\mbox{\epsfxsize 10.0cm\epsfbox{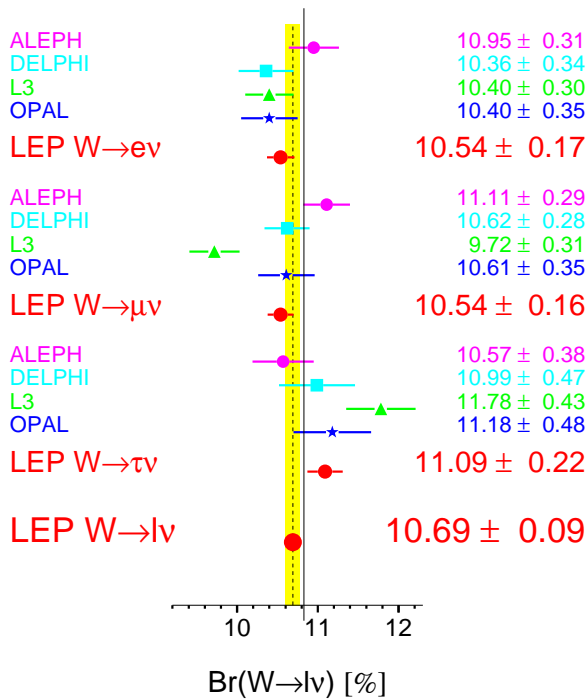}}
\vspace{-3.0cm}
\caption
{LEP results on W leptonic Branching Ratios.}
\label{fig:w}
\end{figure}

Similar tests can be performed with the comparison of the decay widths of the $W$
to the different leptons, $\Gamma(W \to l \nu) \propto g_l^2$. The 
preliminary analysis of the 40000 
W pairs selected at LEP gives the Branching Ratios summarised in
figure~\ref{fig:w}~\cite{lepeew}.
They yield the following ratios of the leptonic couplings:
\begin{flushleft}
$g_\mu/g_e=1.000\pm0.011$\\
$g_\tau/g_\mu=1.026\pm0.014$\\
$g_\tau/g_e=1.026\pm0.014$\\
$g_\tau/g_l=1.026\pm0.010$ (e-$\mu$ univ. assumed)\\
\end{flushleft}
improving significantly  the current Tevatron measurements~\cite{pdg}:
\begin{flushleft}
$g_\mu/g_e=0.986\pm0.029$\\
$g_\tau/g_e=0.988\pm0.025$\\
$g_\tau/g_\mu=1.002\pm0.038$ (not ind. from the above)\\
$g_\tau/g_l=0.988\pm0.025$ (e-$\mu$ univ. assumed)\\
\end{flushleft}
The combination of both sets of measurements give:
\begin{flushleft}
$g_\mu/g_e=0.998\pm0.010$\\
$g_\tau/g_l=1.021\pm0.009$ (e-$\mu$ univ assumed)\\
\end{flushleft}
which are significantly less precise than those from the tau decays, but are
sensitive to different potential new physics, because the intervening W is
`on-shell' and the $q^2$ is much higher. It is interesting to note that
there is an intriguing  hint of a departure from universality
between the $\tau$ 
and light lepton couplings, a discrepancy 
at 2.3 standard deviations. However, this discrepancy is clearly compatible with
a statistical fluctuation and it is still to be confirmed with the final publication
of the results. 

\subsection{Summary on charged current universality and implications}
\begin{figure}[htb]
\begin{center}
\mbox{\epsfxsize 8.5cm\epsfbox{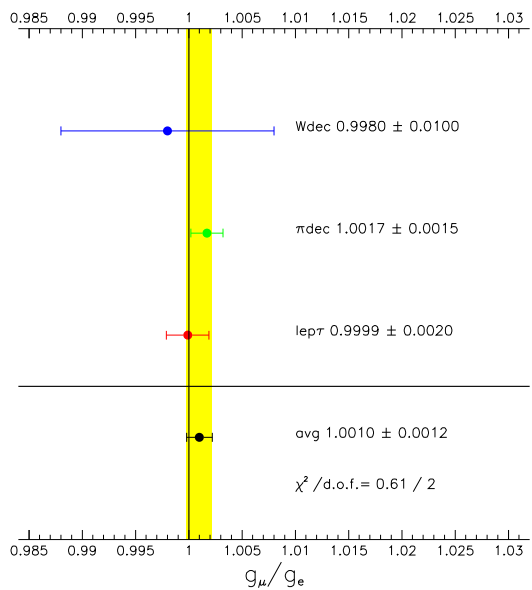}}
\end{center}
\vspace{-1.0cm}
\caption
{Summary of the status of muon-electron Universality measurements with different methods: W decay, pion decay and $\tau$ leptonic
decays.}
\label{fig:univemu}
\end{figure}

\begin{figure}[htb]
\begin{center}
\mbox{\epsfxsize 8.5cm\epsfbox{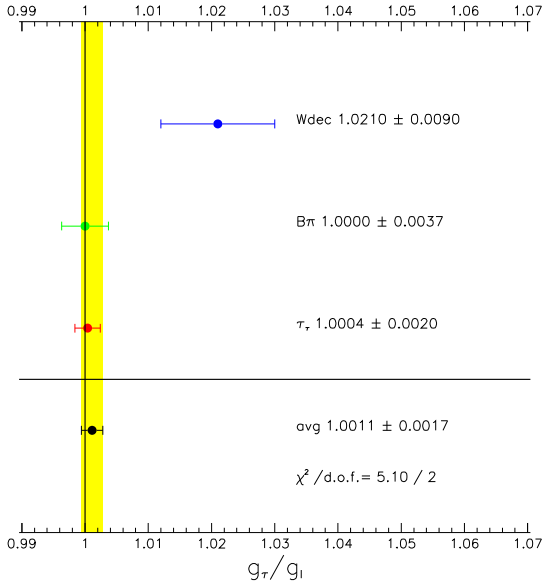}}
\end{center}
\vspace{-1.0cm}
\caption
{Summary of the status of tau-light lepton Universality measurements with different methods: 
W decay, tau decay to pion and $\tau$ lifetime.}
\label{fig:univtau}
\end{figure}

The current state of the art is summarised in figures~\ref{fig:univemu} and~\ref{fig:univtau}. 
These results can be translated into limits on new physics. The limits on a charged higgs are
discussed in the next section in combination with the Michel parameters. Here I will just mention
an example of bounding the tau neutrino mass and the mixing with an
hypothetical $4^{th}$ family neutrino. If the $\nu_\tau$ mass is different from 0 or a $4^{th}$ 
family neutrino with non-zero mixing exist, the different decay amplitudes will be affected
~\cite{boundsswain}, with the consequent reflect on the universality of the couplings. The
current measurements allow to set the following limits at 95\% C.L.:
\begin{flushleft}
$m_{\nu\tau}<32$ MeV\\
$|\sin\theta_{\nu\tau-\nu 4}|<0.057$\\
\end{flushleft}

\subsection{Lorentz structure}
In the Standard Model the charged current interaction is assumed to
be of the type $V$-$A$, a vector and an axial-vector couplings with
the same magnitude and opposite sign. However, there is no fundamental
reason for that and the existence of more general couplings is still possible. 
There are
stringent experimental results that this assumption is correct on the muon decay.
Assuming the lepton number conservation, derivative free, local and Lorentz
invariant 4-fermion point interaction, the most general form for the amplitude
of the tau (muon) decay involves 12 complex couplings~\cite{michel} two of which
must be exactly 0. The tau decay products energy spectra can be expressed in this general form
in terms of five parameters, the $\nu_\tau$ helicity and 
the so called `Michel parameters', $\eta$, $\rho$, $\xi$ and $\delta$,
in addition of the momentum and $\ptau$. 
Therefore, these parameters are experimentally accessible from these
distributions and as a consequence information on the couplings can be inferred.

Of all these parameters, only $\eta$ affects the partial widths, while the remaining ones distort
the differential cross section but do not change the normalization. 
In particular, assuming that the experiments have a leptonic selection 
whose efficiency does not depend on the momentum and with an infinite momentum resolution
$BR(\tau \to l \nu \nu) \propto (1+\alpha \eta  \frac{m_l}{m_\tau})$, with $\alpha=4$. 
Then the ratio of the Branching
Ratios provides a measurement of $\eta$. However, experimental effects make $\alpha<4$, 
but often the experiments do not
calculate (or do not quote) the precise value for their particular conditions (see~\cite{stugu} for a thorough
discussion on the subject). 

\begin{figure}[htbp]
\begin{center}
\mbox{\epsfxsize 8.0cm\epsfbox{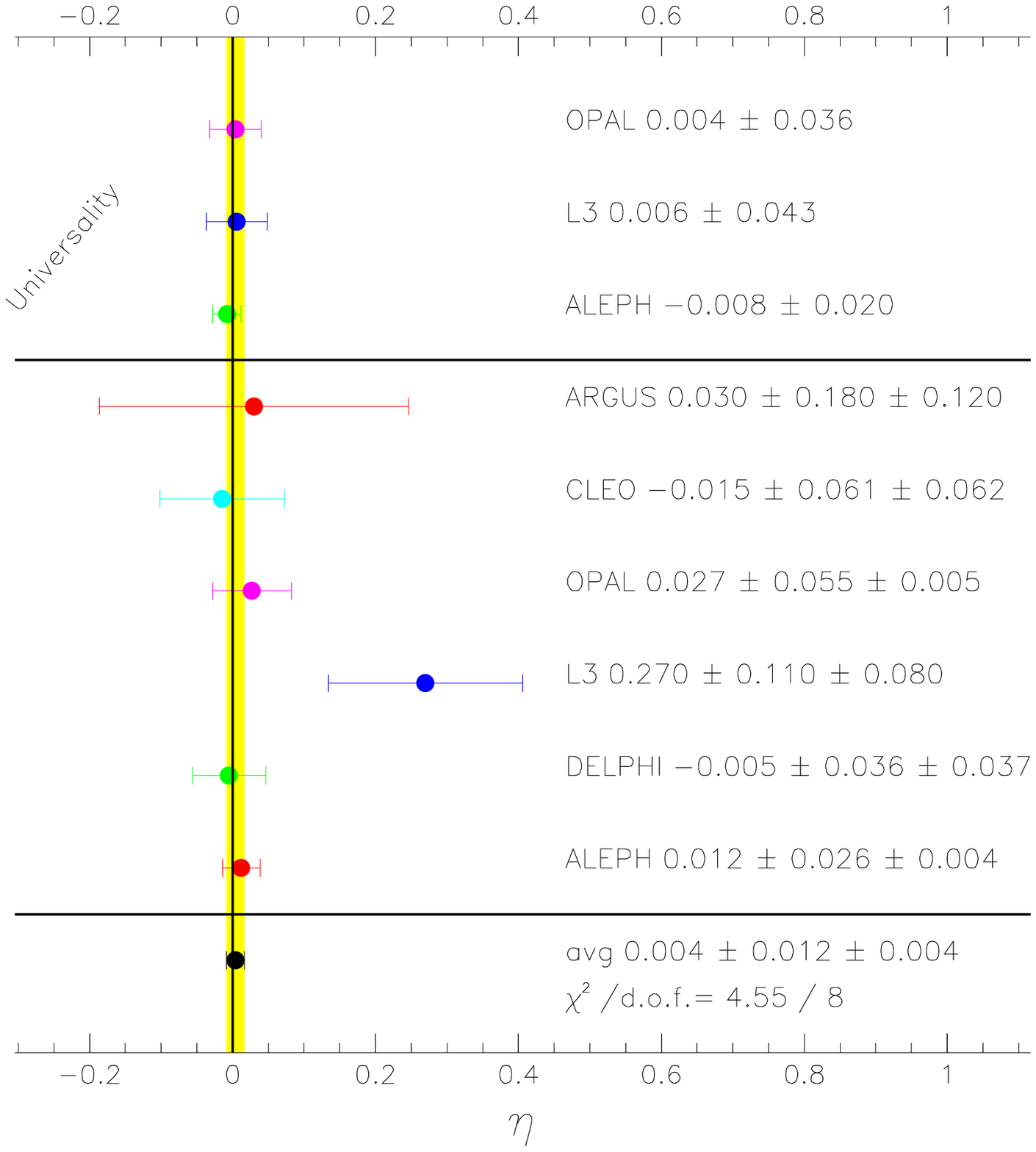}}
\mbox{\epsfxsize 8.0cm\epsfbox{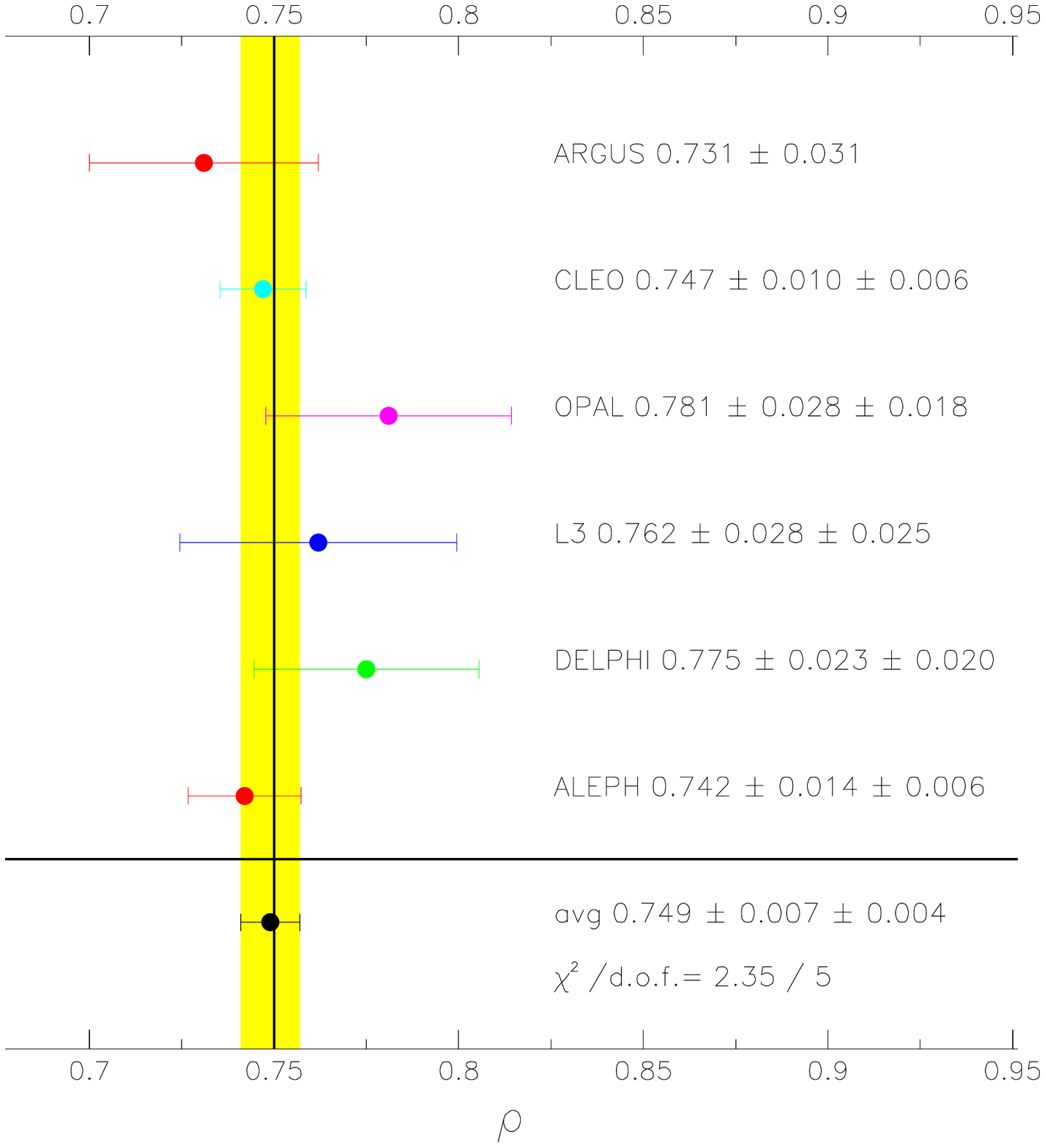}}
\end{center}
\vspace{-1.0cm}
\caption
{Results on the Michel parameters $\eta$ and $\rho$. 
The vertical line shows the Standard Model expectation for $V$-$A$ structure in the charged current
current.}
\label{fig:michel1}
\end{figure}

\begin{figure}[htbp]
\begin{center}
\mbox{\epsfxsize 8.0cm\epsfbox{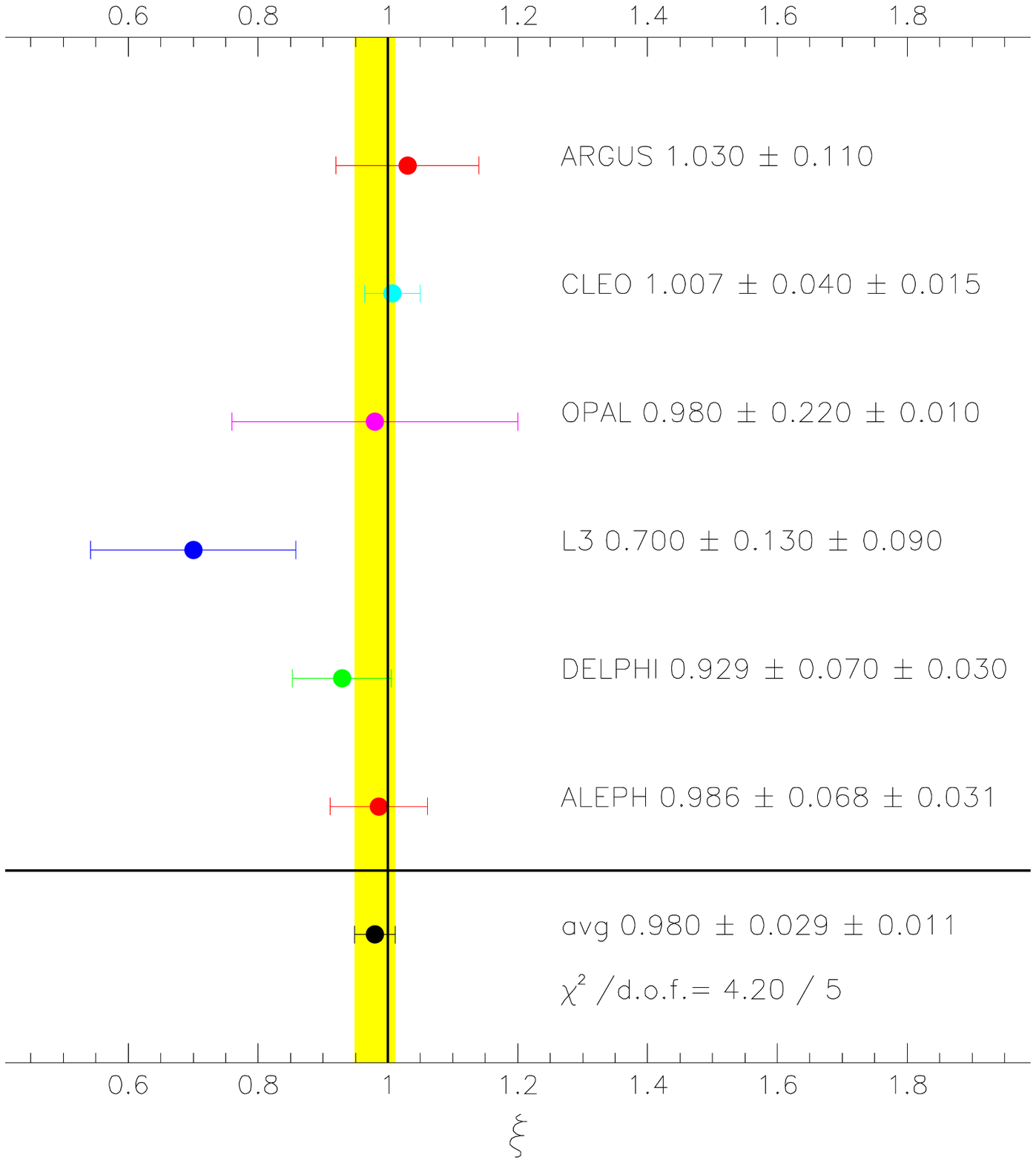}}
\mbox{\epsfxsize 8.0cm\epsfbox{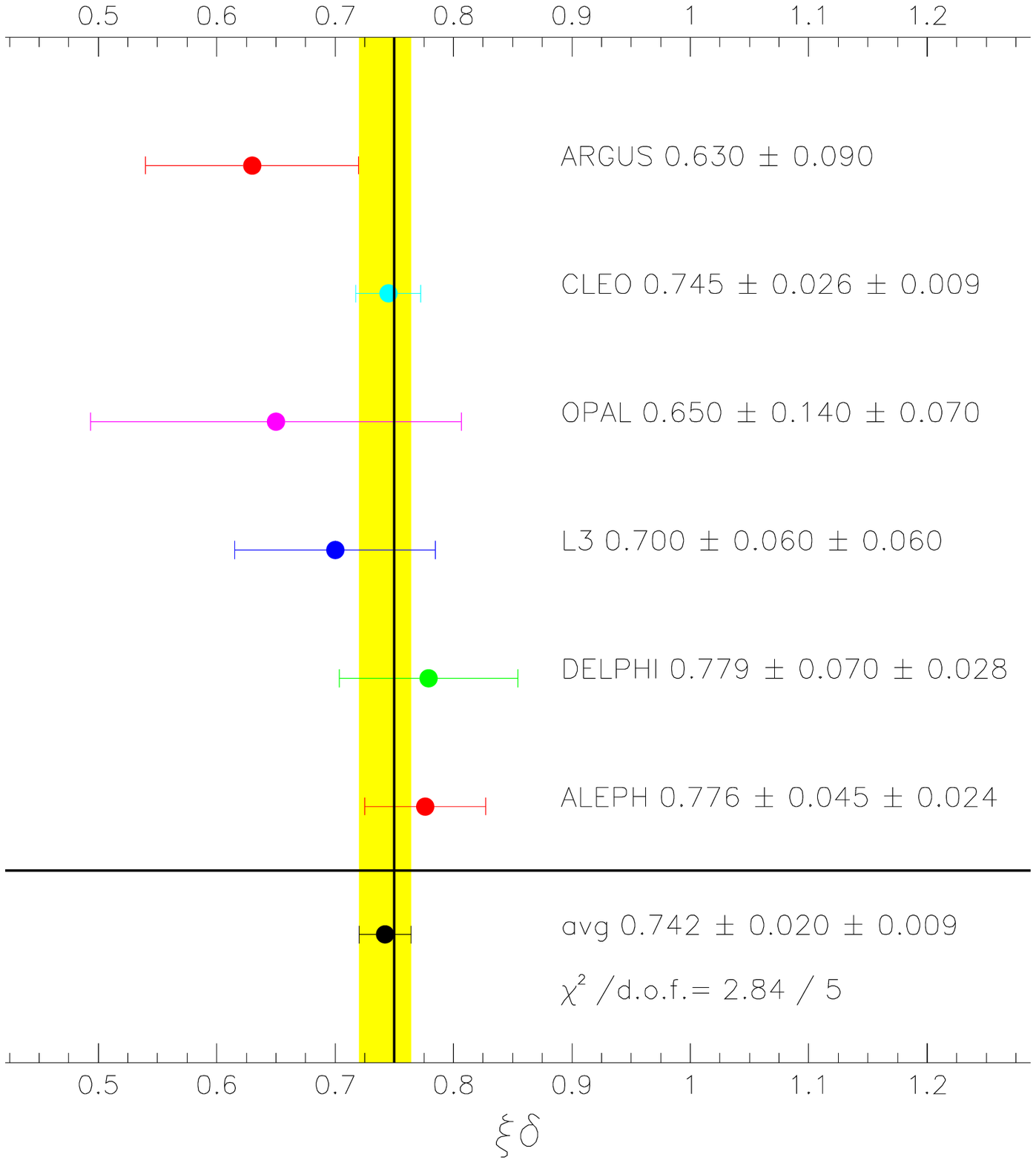}}
\end{center}
\vspace{-1.0cm}
\caption
{Results on the Michel parameters $\xi$ and $\xi\delta$. 
The vertical line shows the Standard Model expectation for $V$-$A$ structure in the charged current
current.}
\label{fig:michel2}
\end{figure}

\begin{figure}[htbp]
\begin{center}
\mbox{\epsfxsize 8.0cm\epsfbox{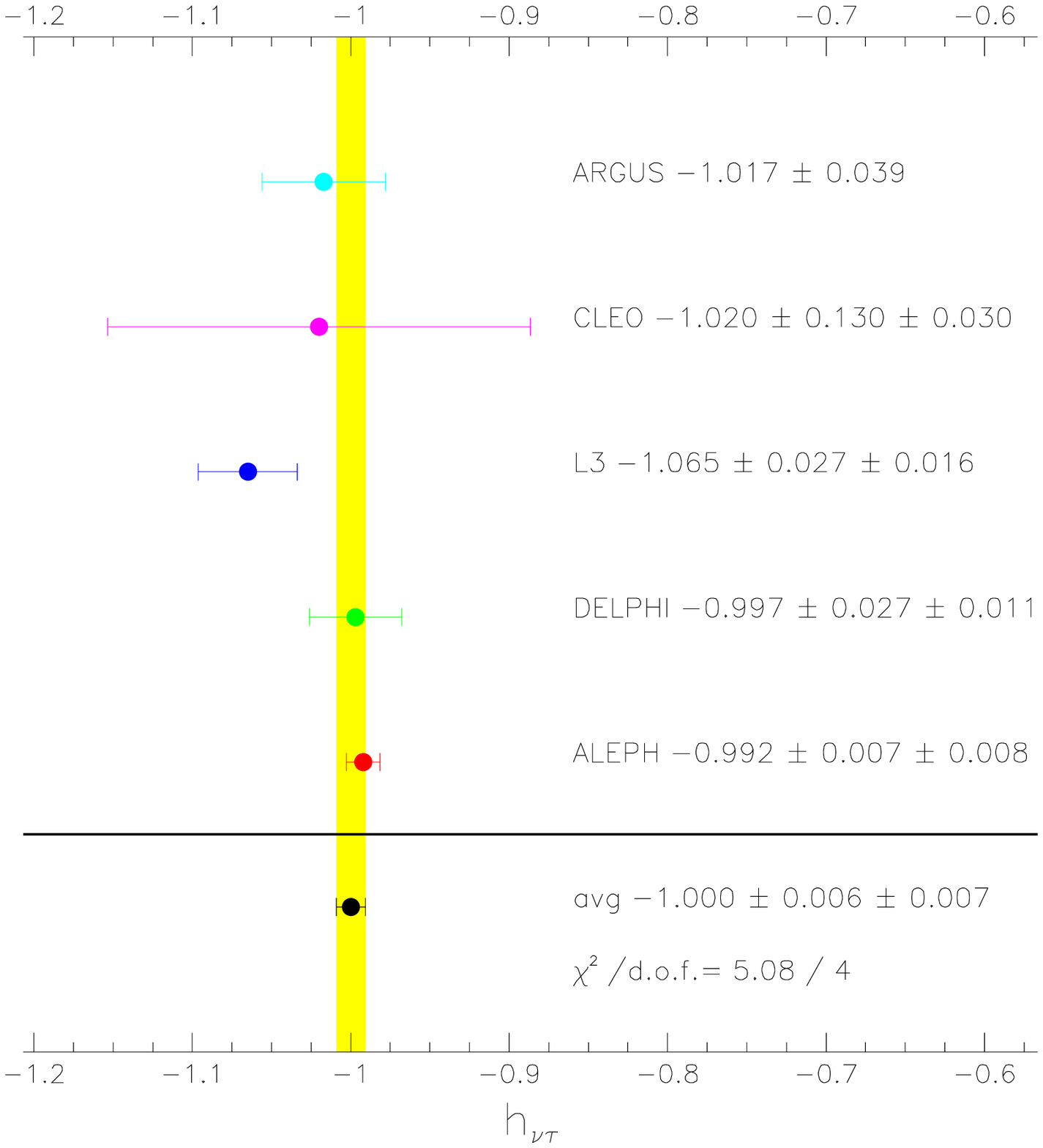}}
\end{center}
\vspace{-1.0cm}
\caption
{Results on the $\nu_\tau$ helicity. 
The vertical line shows the Standard Model expectation for $V$-$A$ structure in the charged current
current.}
\label{fig:heli}
\end{figure}

Figures~\ref{fig:michel1} to~\ref{fig:heli} show the final results from LEP
experiments~\cite{alephlor,delphilor,l3lor,opallor}, compared to other recent
results~\cite{pdg} and with the Standard Model
expectation for a pure $V$-$A$ structure, on the assumption of lepton universality for the parameters. 
ALEPH, DELPHI
and OPAL have also measured the parameters on the assumption that they might be different for
electrons and muons, not showing any deviations. DELPHI includes in the fit a constraint from
universality, with the correct $\alpha$. For the remaining experiments, the
$\eta$ is also shown as obtained from the universality on the assumption that $\alpha=4$ (known to be a
reasonably good approximation for LEP experiments~\cite{stugu}).

DELPHI has studied~\cite{delphilor} an additional anomalous tensor coupling predicted by a model in
which the Lagrangian containing derivatives~\cite{tensor}. In a similar manner this term distorts the
momentum distributions of the decays products and its strength is measurable. This strength has been bounded to
$\kappa<0.050$ (95\% CL).

\subsection{Bounds on new physics}

The precision tests on the universality and on the Lorentz structure, set bounds on different extensions of
the Standard Model. 

The existence of a charged Higgs would influence the Michel parameters. Within MSSM, the
$\eta$ parameter would have a non-zero value of $\eta_l\approx \frac{m_l m_\tau}{2 M_H^2} {\tan \beta}^2 $, while
the rest of the parameters would remain unchanged or with a variation of 
second order on $\frac{m_l m_\tau}{2 M_H^2}$.  
From the previous results  we can set the limit at 95\% CL: $M_{H^\pm}>2.4 \tan (\beta)$, which is only
competitive with direct searches for high $\tan(\beta)$.

DELPHI~\cite{delphilor} and OPAL~\cite{opallor} have also interpreted their results in the light of the
possible existence of an additional vector boson. 
DELPHI sets the limit at 95\% CL: $M_{W_2} > 189~ GeV$ for any mixing with the
standard $W$ or $-0.141<\eta<0.125$ rad for the mixing and any boson mass.
OPAL sets the limit at 95\% CL: $M_{W_2} > 137~ GeV$ for any mixing with the
standard $W$ or $|\eta|<0.12$ rad for the mixing and any boson mass.

\section{Hadronic Branching Ratios}

\begin{figure}[htbp]
\begin{center}
\mbox{\epsfxsize 8.0cm\epsfbox{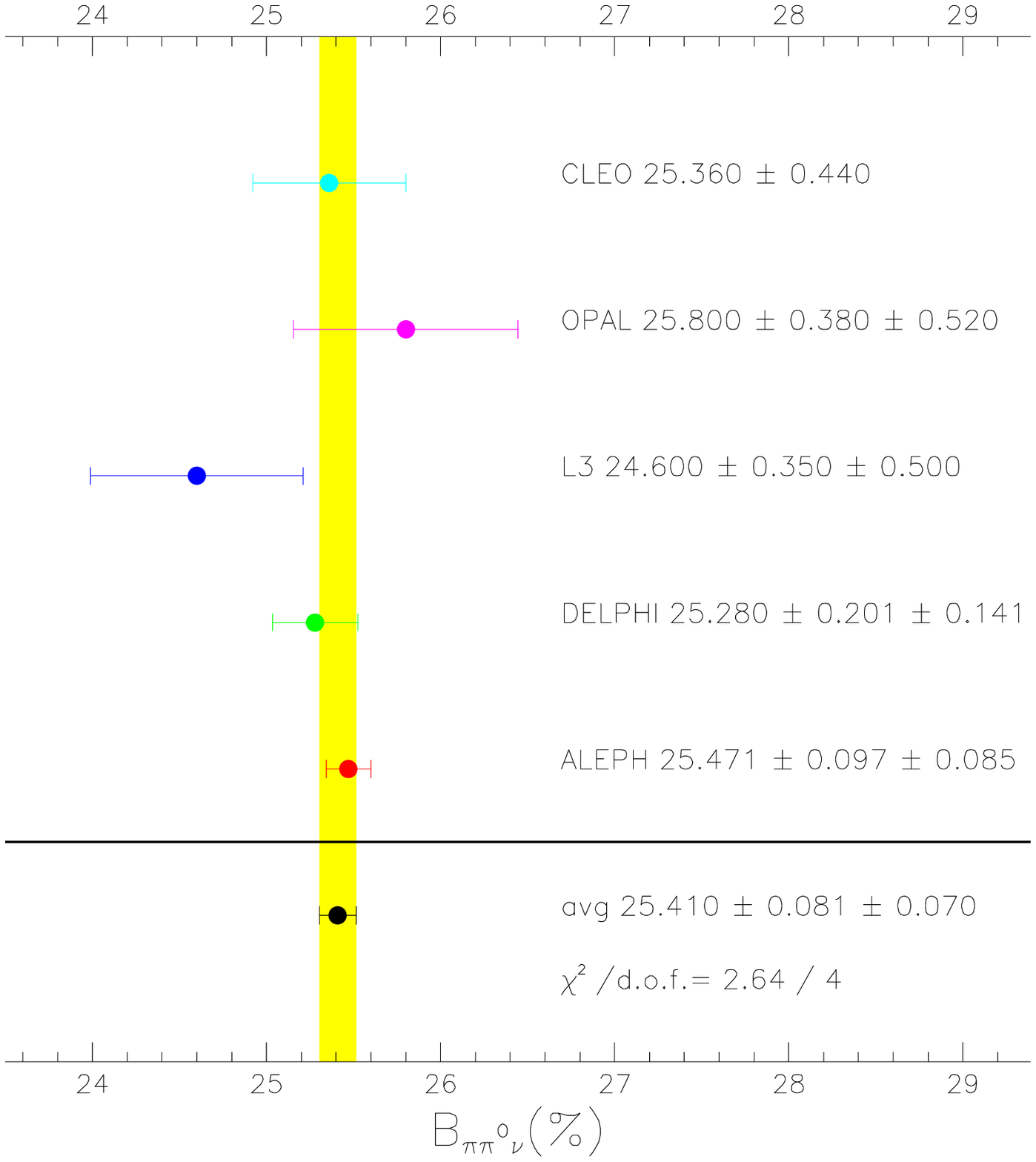}}
\end{center}
\vspace{-1.0cm}
\caption{Branching Ratio for the decay $\TRO$.}
\label{fig:bpipi0}
\end{figure}

\begin{figure}[htbp]
\begin{center}
\mbox{\epsfxsize 8.0cm\epsfbox{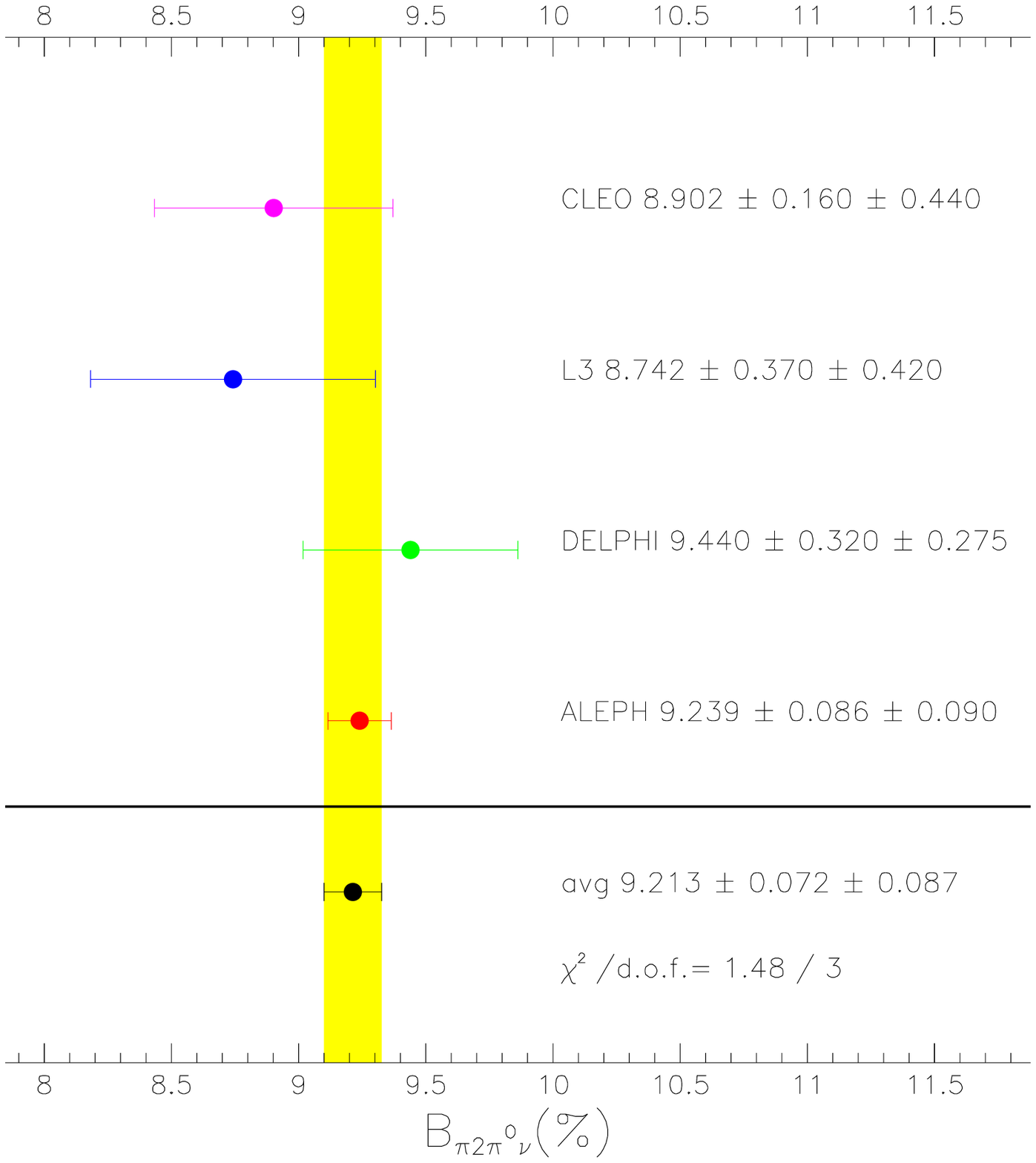}}
\end{center}
\vspace{-1.0cm}
\caption{Branching Ratio for the decay $\TAA$.}
\label{fig:bpi2pi0}
\end{figure}

\begin{figure}[htbp]
\begin{center}
\mbox{\epsfxsize 8.0cm\epsfbox{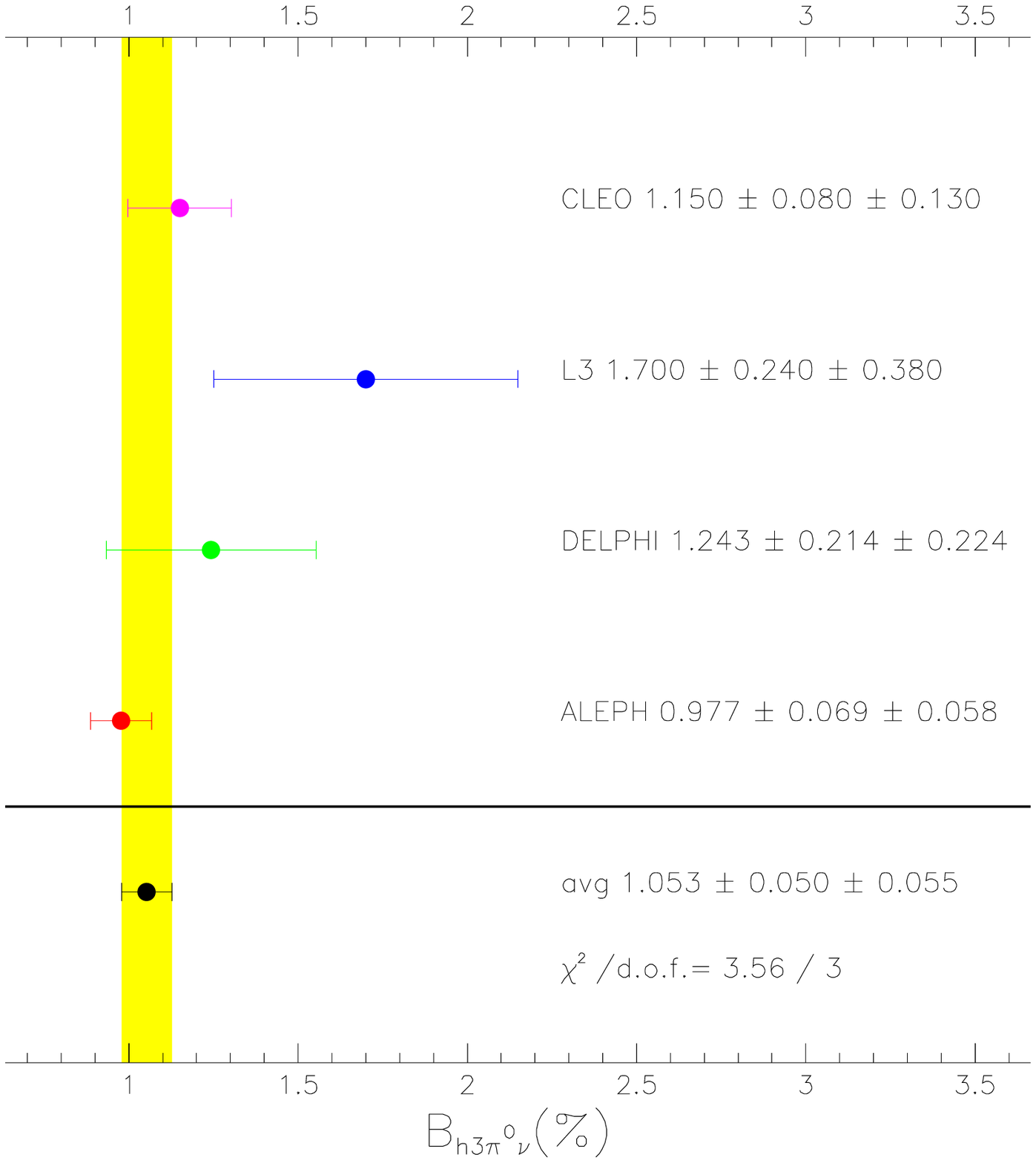}}
\end{center}
\vspace{-1.0cm}
\caption{Branching Ratio for the decay $\TPMP$.}
\label{fig:bpi3pi0}
\end{figure}

\begin{figure}[htbp]
\begin{center}
\mbox{\epsfxsize 8.0cm\epsfbox{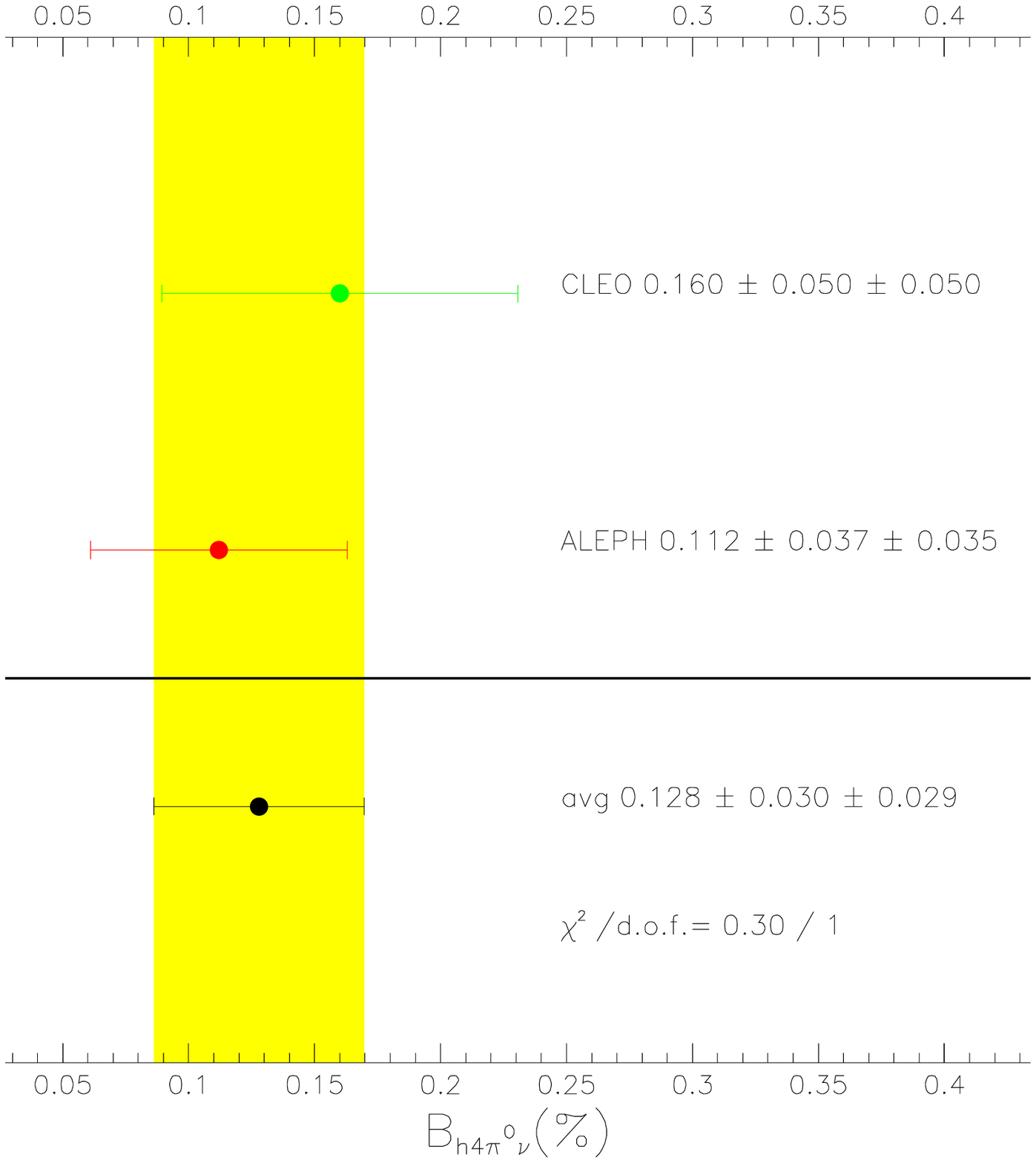}}
\end{center}
\vspace{-1.0cm}
\caption{Branching Ratio for the decay $\TPMM$.}
\label{fig:bpi4pi0}
\end{figure}

\begin{figure}[htbp]
\begin{center}
\mbox{\epsfxsize 8.0cm\epsfbox{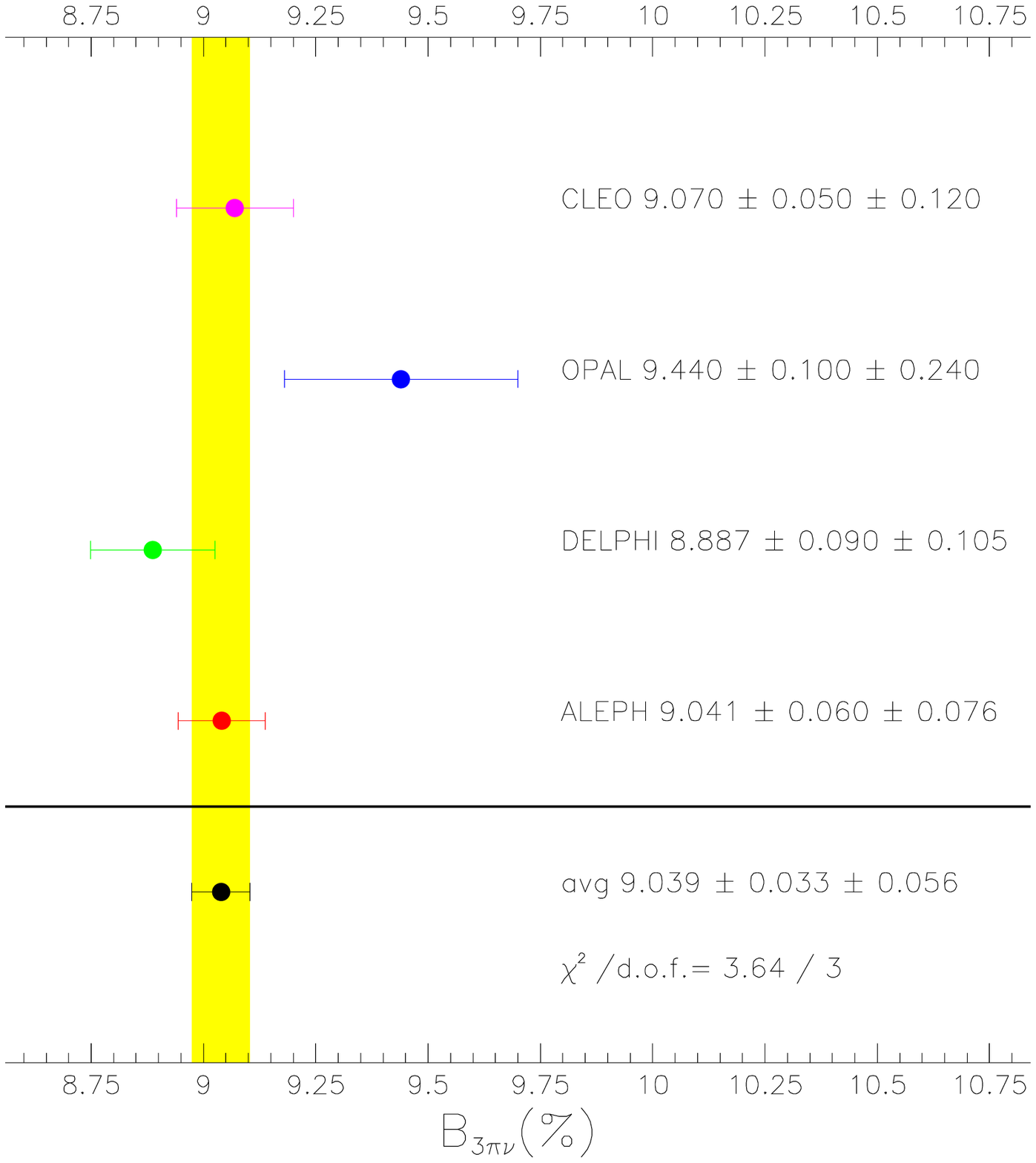}}
\end{center}
\vspace{-1.0cm}
\caption{Branching Ratio for the decay $\TPPP$.}
\label{fig:b3pi}
\end{figure}

\begin{figure}[htbp]
\begin{center}
\mbox{\epsfxsize 8.0cm\epsfbox{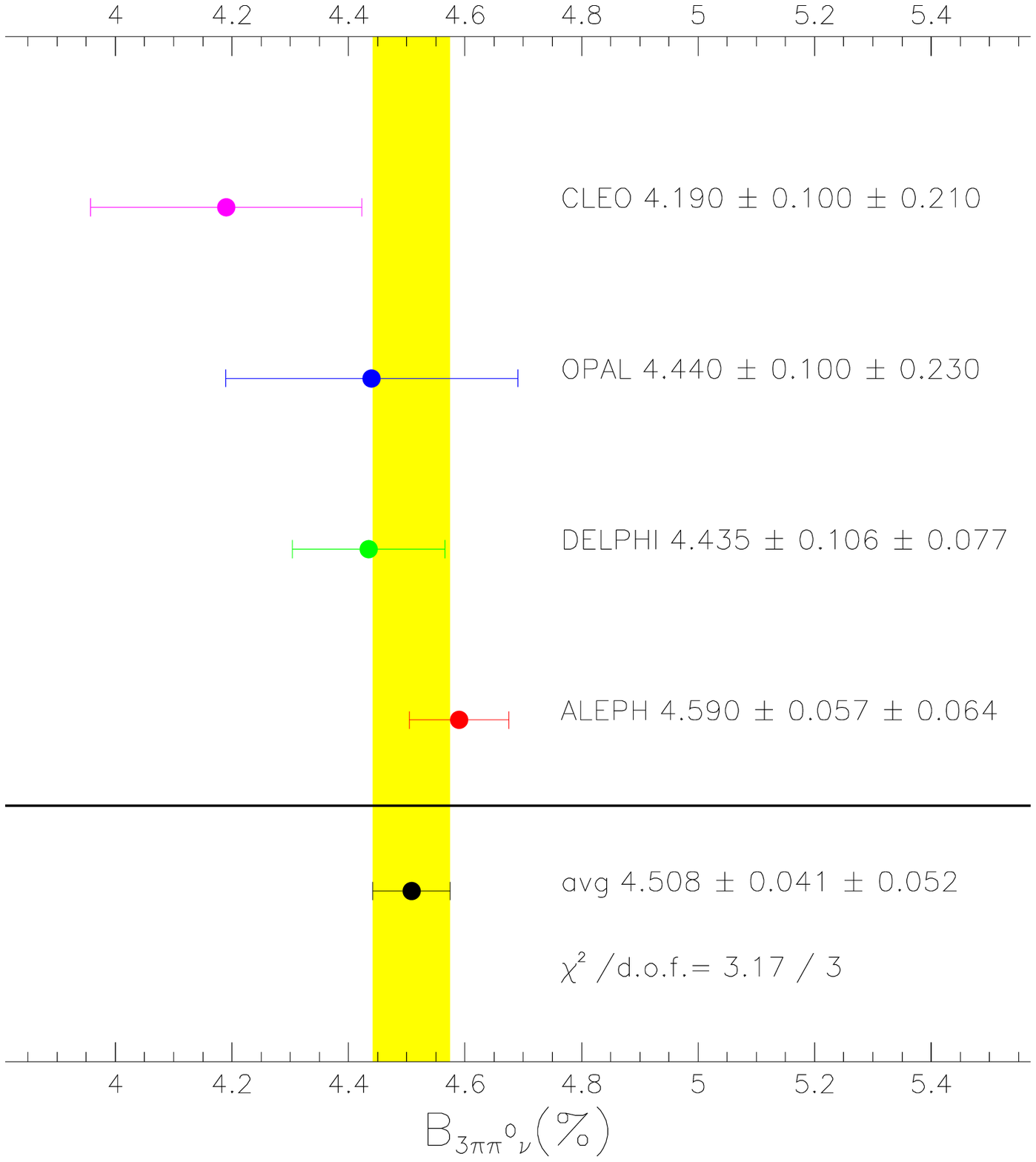}}
\end{center}
\vspace{-1.0cm}
\caption{Branching Ratio for the decay $\TPPPO$.}
\label{fig:b3pipi0}
\end{figure}

\begin{figure}[htbp]
\begin{center}
\mbox{\epsfxsize 8.0cm\epsfbox{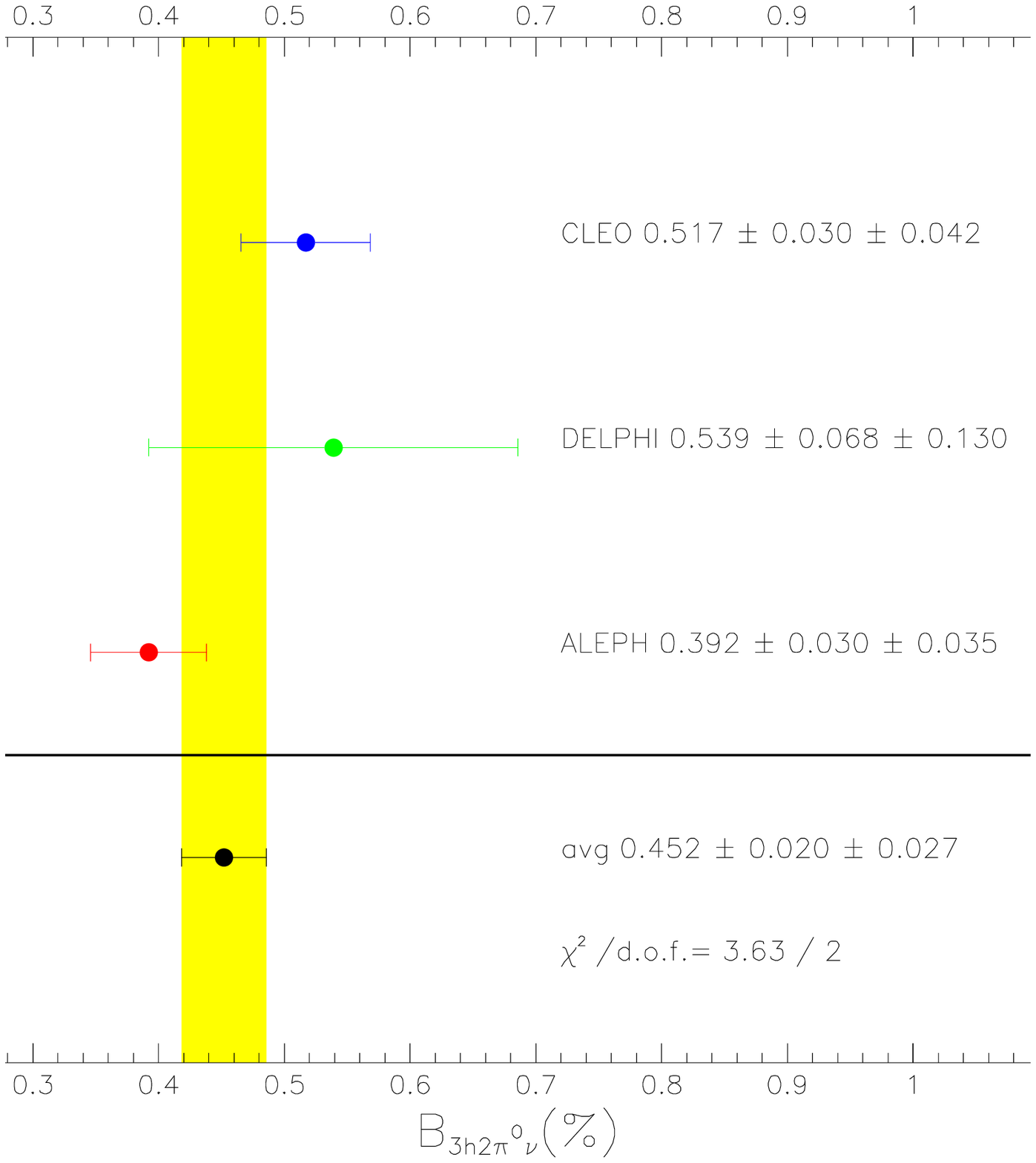}}
\end{center}
\vspace{-1.0cm}
\caption{Branching Ratio for the decay $\TPPPOO$.}
\label{fig:b3pi2pi0}
\end{figure}

\begin{figure}[htbp]
\begin{center}
\mbox{\epsfxsize 8.0cm\epsfbox{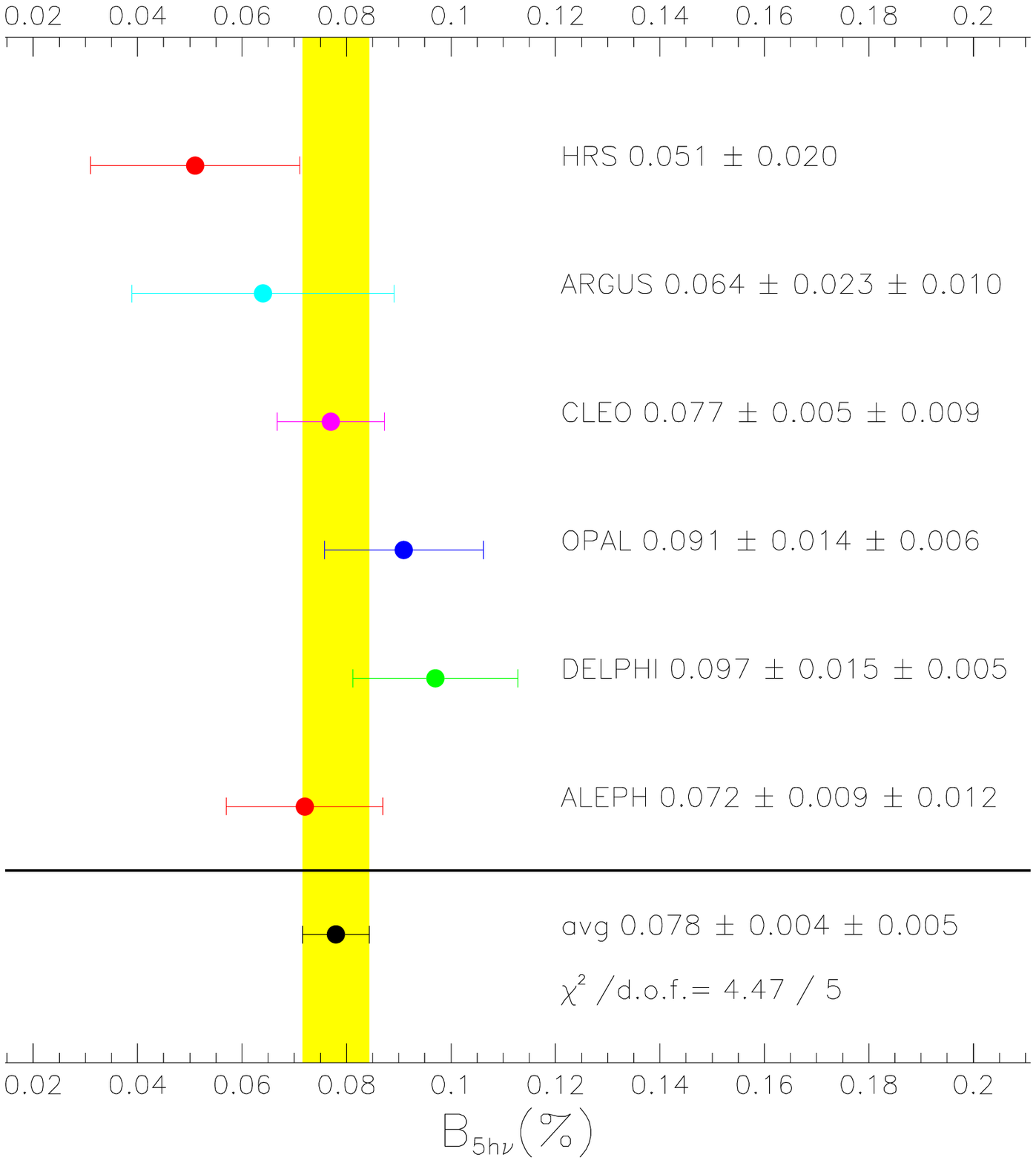}}
\end{center}
\vspace{-1.0cm}
\caption{Branching Ratio for the decay $\TPPPPP$.}
\label{fig:b5pi}
\end{figure}

\begin{figure}[htbp]
\begin{center}
\mbox{\epsfxsize 8.0cm\epsfbox{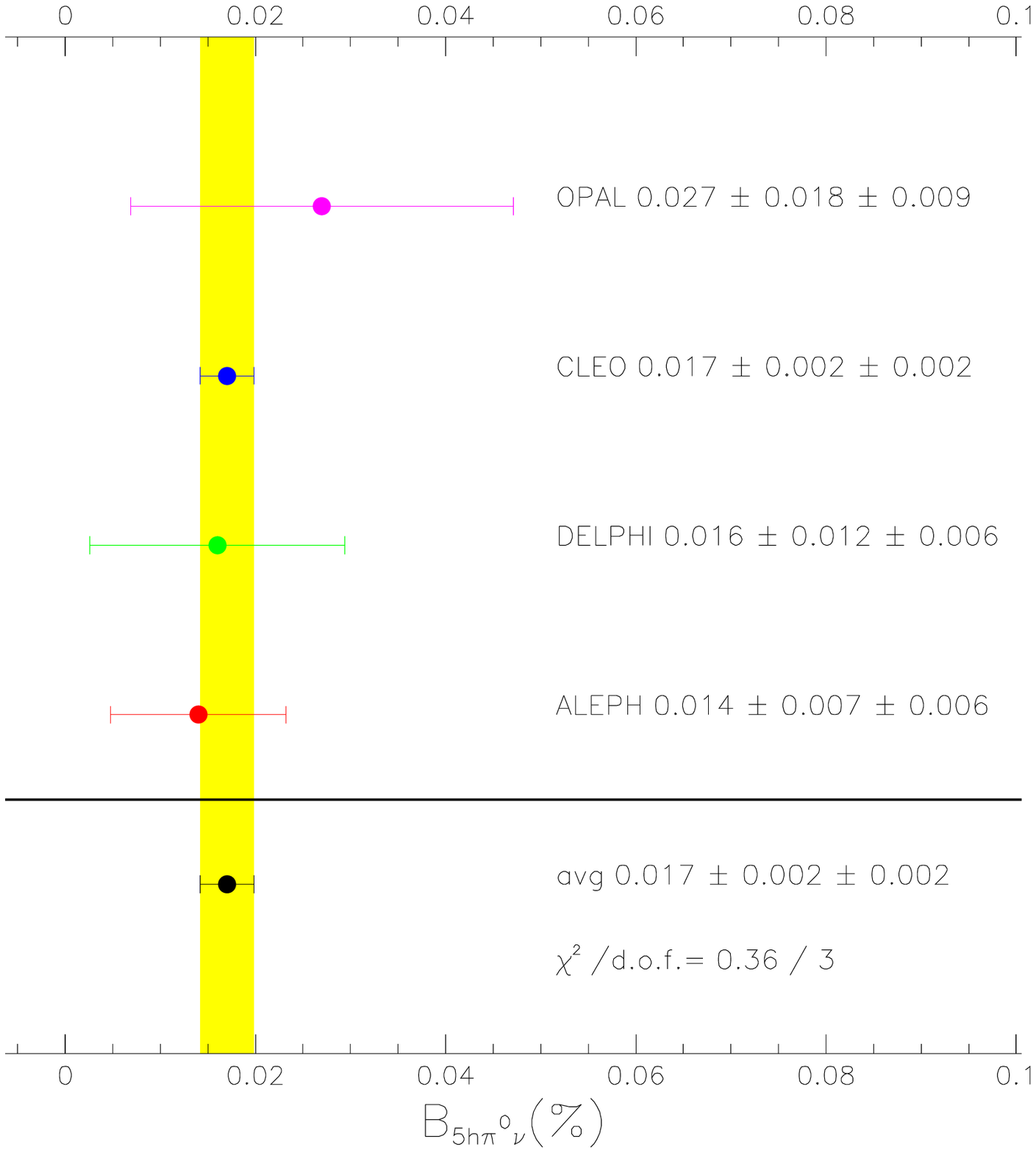}}
\end{center}
\vspace{-1.0cm}
\caption{Branching Ratio for the decay $\TPPPPPO$.}
\label{fig:b5pipi0}
\end{figure}

In addition to the previously mentioned $\TPI$ decay, ALEPH~\cite{davier} and DELPHI~\cite{delphihad} have
recently presented new preliminary results for the Branching Ratios not involving kaon identification 
for a large variety of hadronic decays (with up to 6 neutral or charged hadrons). Figures~\ref{fig:bpipi0} to ~\ref{fig:b5pipi0}, summarise these
results, together with other existing precise measurements~\cite{pdg}. When possible, the results are quoted with the
kaon component subtracted. For ALEPH and some CLEO measurements, this subtraction is done by the same
experiment using their own data, while for the remaining cases the w.a.~\cite{pdg} were used.

\begin{figure}[htbp]
\begin{center}
\mbox{\epsfxsize 8.0cm\epsfbox{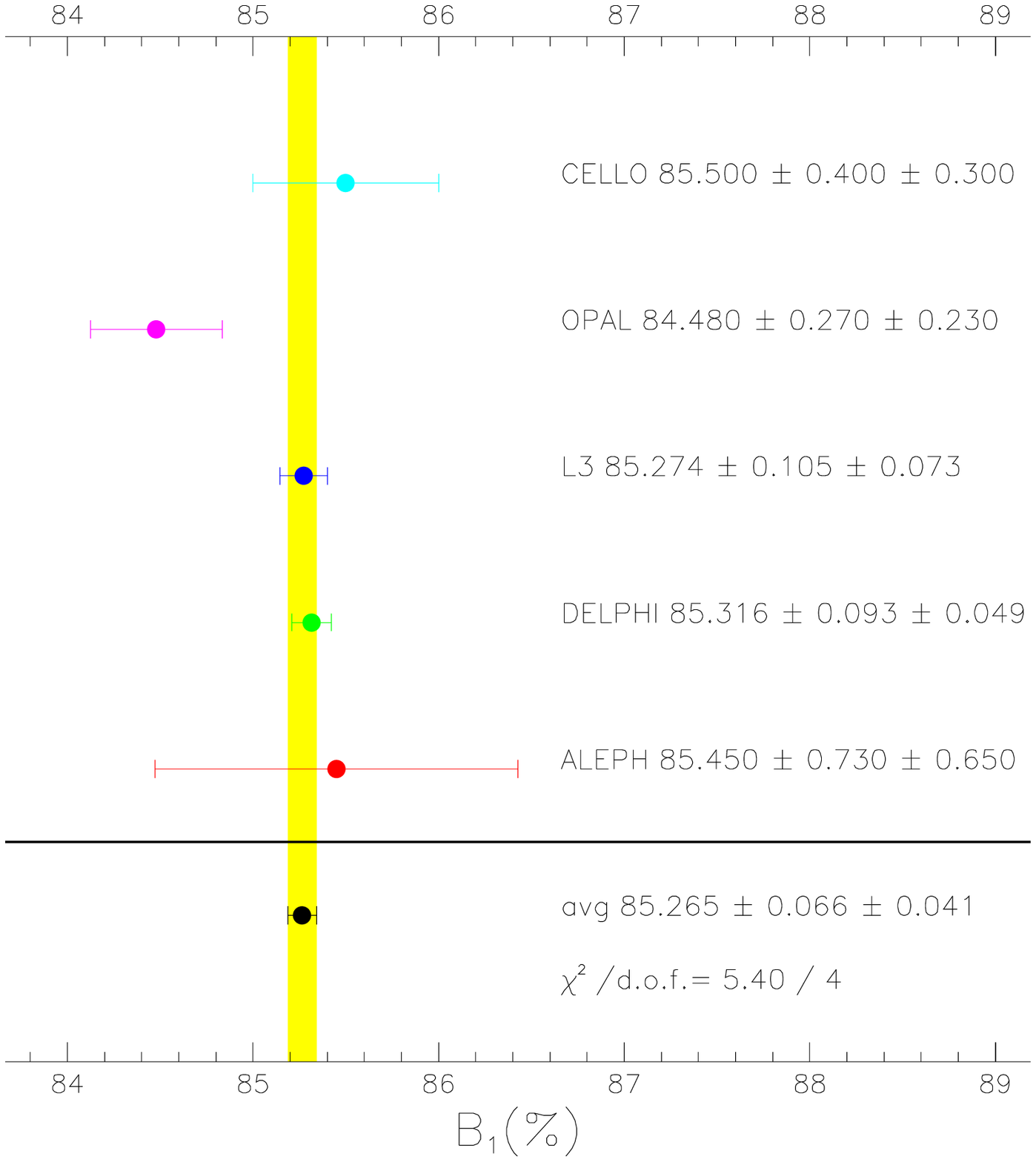}}
\end{center}
\vspace{-1.0cm}
\caption{Branching Ratio to one charged particle (1-prong).}
\label{fig:b1}
\end{figure}

\begin{figure}[htbp]
\begin{center}
\mbox{\epsfxsize 8.0cm\epsfbox{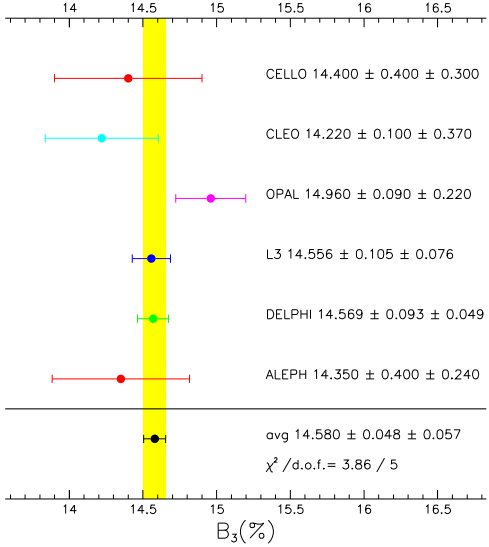}}
\end{center}
\vspace{-1.0cm}
\caption{Branching Ratio to three charged particles (3-prongs).}
\label{fig:b3}
\end{figure}

\begin{figure}[htbp]
\begin{center}
\mbox{\epsfxsize 8.0cm\epsfbox{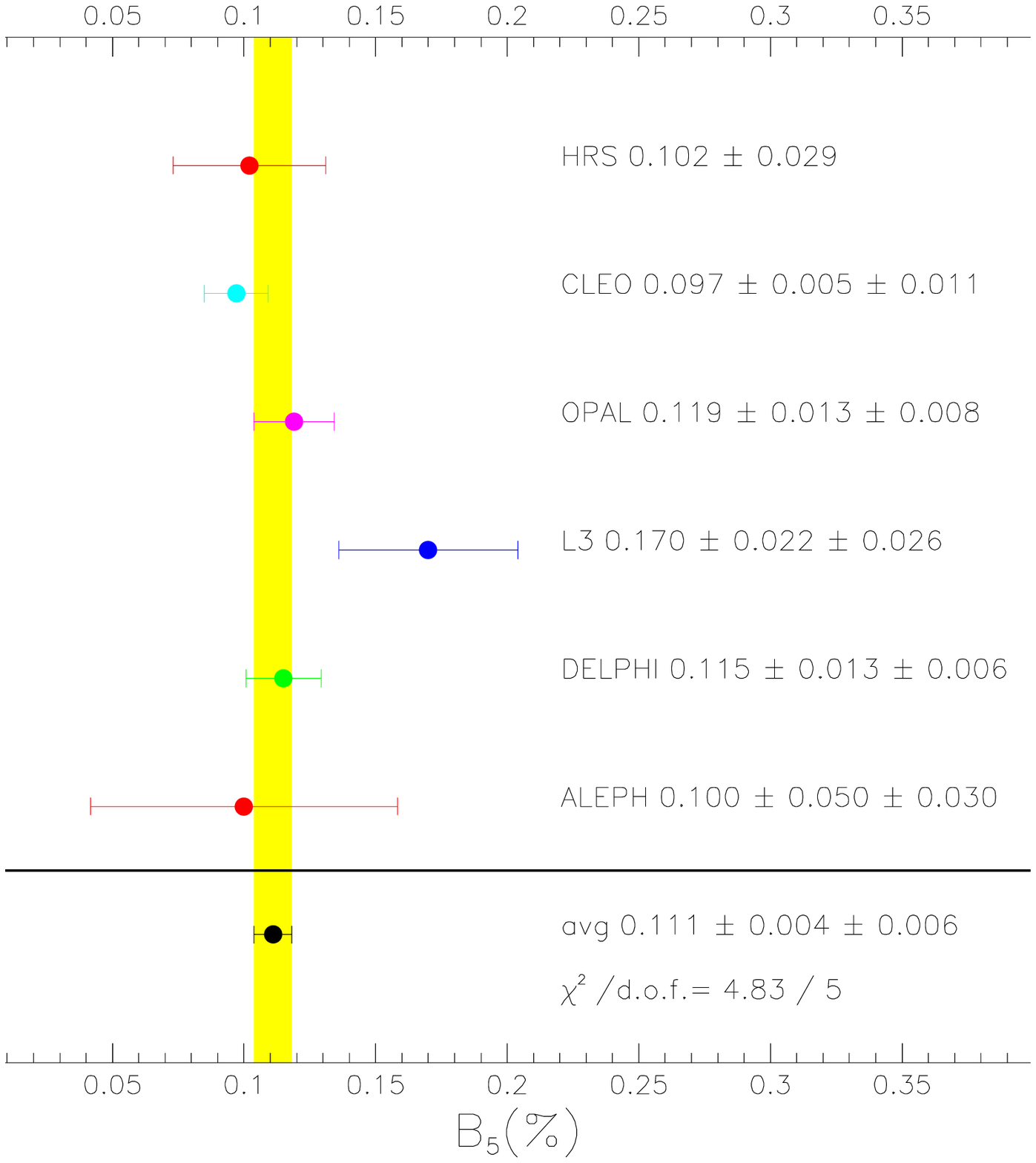}}
\end{center}
\vspace{-1.0cm}
\caption{Branching Ratio to five charged particles (5-prongs).}
\label{fig:b5}
\end{figure}

DELPHI~\cite{delphitopo} and L3~\cite{l3topo} have also published new measurements on the topological Branching Ratios.
The results are shown in figures~\ref{fig:b1} to ~\ref{fig:b5}. In these measurements, the $\ko$ are considered as neutral
particles, regardless of its decay, and therefore their possible decays products were not counted as "primary charged tracks" in
the definition. These new results basically supersede all the previous results for one and three prongs, 
solving some remaining inconsistencies.

As an interesting check of the completeness of the $\tau$ decays, or the consistency of the measurements, we can show the
perfect agreement between the 
current average for the direct measurement of the Branching Fraction to one-prong, $85.265\pm0.078 \%$ with the sum of all the
exclusive modes contributing to this topology as measured by ALEPH, $85.265\pm0.110 \%$. It is important to remark that these two
numbers are totally independent experimentally and that ALEPH result is the most precise result for this sum, obtained from a
single experiment and fully independent from the topologic result and accounting for all correlations ($DELPHI$ results are
strongly correlated with the topological BR). This comparison shows that there is no hint of the "missing one-prong problem" as
observed in the eighties, nor of some remaining discrepancies on the late nineties. 

\section{Other topics in $\tau$ physics}

\subsection{Lepton flavour violation}

The possible violation of the lepton flavour conservation in the weak neutral current was studied looking for
Z decays to $e\mu$, $e\tau$ or $\mu\tau$. This measurement benefited from the high statistics of Z available
at LEP. The following bounds were set at $95\%$ CL:
\begin{eqnarray*}
BR(Z\to e\mu)<1.7~10^{-6}\\
BR(Z\to e\tau)<9.8~10^{-6}\\
BR(Z\to \mu\tau)<1.2~10^{-6}\\
\end{eqnarray*}

\subsection{$\tau$ mass}
\begin{figure}[htb]
\vspace{0.1cm}
\begin{center}
\mbox{\epsfxsize 6.5cm\epsfysize 5.0cm\epsfbox{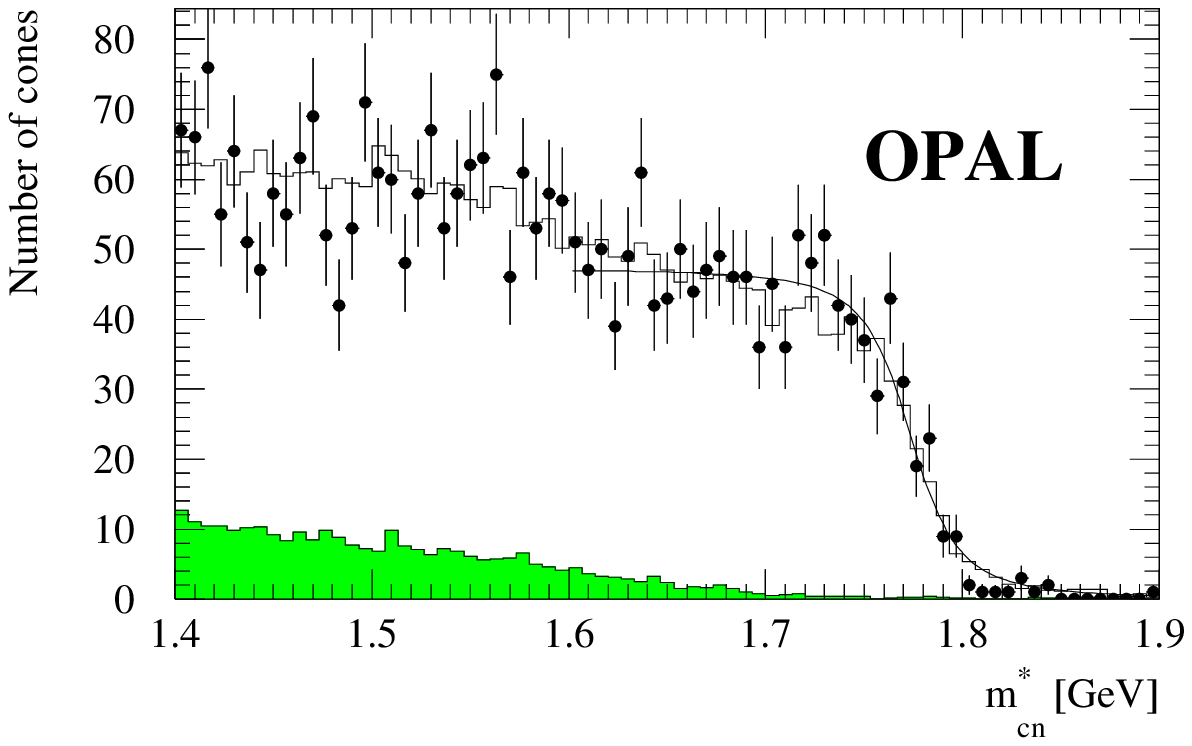}}
\mbox{\epsfxsize 6.5cm\epsfbox{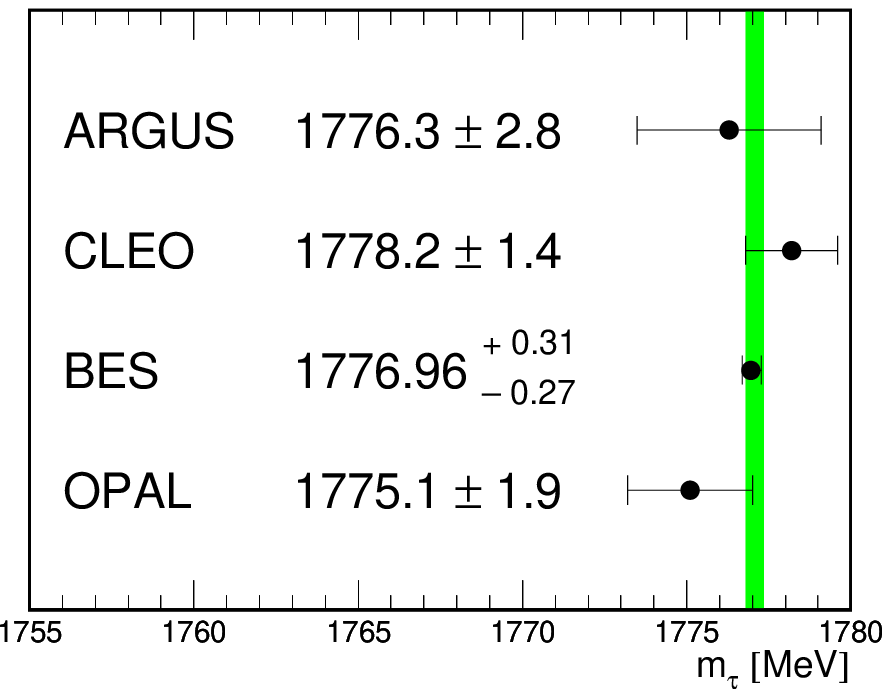}}
\end{center}
\vspace{-1.0cm}
\caption
{Pseudomass distribution used to obtain $m_\tau$ (top) and comparison of current results on $m_\tau$ (bottom).}
\label{fig:mas}
\end{figure}

OPAL has measured the $\tau$ mass at LEP~\cite{taumass} using a pseudomass estimator, and fitting the
mass from the endpoint of this distribution (figure~\ref{fig:mas}). The same figure also shows the comparison with other
experiments. More interesting is that this method can be applied
separately to the $\tau^+$ and $\tau^-$, providing for the first time some information on the difference
between the mass of positive and negative tau and therefore checking CPT invariance. A limit of
$\frac{\Delta(M)}{M}<0.3\%$ was set at 90\% CL. 

\section{Conclusions}
The study of the $\tau$ lepton production and decay at LEP has provided a lot of Standard Model precision measurements
both in the electroweak sector, neutral and charged current universality and structure. 
No deviation from the Standard Model has been observed and
therefore indirect bounds on extensions to Standard Model were set.

LEP experiments are also contributing to the improvement on the hadronic decays Branching Ratios measurement.
These, together with
more precise measurements at LEP on topological Branching Ratios rule out the remaining discrepancy in the
`missing one prong problem' at the 0.1\% level.

\end{document}